\documentclass[11pt,a4paper]{article}
\usepackage{jheppub_mod}
\usepackage[utf8]{inputenc}

\usepackage[dvipsnames]{pstricks}
\usepackage{cleveref}
\crefname{appendix}{Appendix}{Aps.}%
\crefname{table}{Table}{Tables}%
\crefname{section}{Sec.}{Secs.}%
\crefname{figure}{Fig.}{Figs.}%
\crefname{equation}{Eq.}{Eqs.}%
\usepackage{subcaption}
\usepackage{booktabs}

\hypersetup{
  pdfauthor={Nienke Balz, Florian Herren, Bastian Kubis, Simon Mutke, Méril Reboud},
  pdftitle={Advanced parametrisations for hadronic form factors},
  pdfkeywords={Form factors, hadronic resonances},
  colorlinks = true
}

\allowdisplaybreaks

\renewcommand{\d}{\mathrm{d}}

\definecolor{darkgreen}{rgb}{0.0, 0.2, 0.13}
\definecolor{darkspringgreen}{rgb}{0.09, 0.45, 0.27}

\title{Advanced parametrisations for hadronic form factors}
\preprint{ZU-TH 70/25}

\author[a]{Nienke C. Balz,}
\author[b]{Florian Herren,}
\author[a]{Bastian Kubis,}
\author[a]{Simon Mutke,}
\author[c]{and Méril Reboud}

\affiliation[a]{Helmholtz-Institut für Strahlen- und Kernphysik (Theorie) and
Bethe Center for Theoretical Physics, Universität Bonn, 53115 Bonn, Germany}
\affiliation[b]{Physics Institute, Universität Zürich, Winterthurerstrasse 190, CH-8057 Zürich, Switzerland}
\affiliation[c]{Université Paris-Saclay, CNRS/IN2P3, IJCLab, 91405 Orsay, France}
\emailAdd{balz@hiskp.uni-bonn.de}
\emailAdd{florian.s.herren@gmail.com}
\emailAdd{kubis@hiskp.uni-bonn.de}
\emailAdd{mutke@hiskp.uni-bonn.de}
\emailAdd{meril.reboud@cnrs.fr}

\abstract{
The rich analytic structure of hadronic form factors makes a theoretically consistent yet easily applicable parametrisation cumbersome.
Consequently, most parametrisations are limited to reproducing the simplest analytic features sufficient to describe form factors on their first Riemann sheet.
Here, we introduce two novel form factor parametrisations that allow resonance poles and left-hand cuts on the second Riemann sheet to be studied, while also making the connection to partial-wave amplitudes manifest.
}

\begin{document}
\maketitle

\section{Introduction}
Production and decays of hadrons are governed by the low-energy dynamics of Quantum Chromodynamics (QCD) and, consequently, fall outside the range of perturbation theory.
However, Lorentz invariance of the relevant amplitudes allows for a decomposition into tensor structures and multi-valued scalar functions, known as form factors (FFs).
To obtain predictions, these FFs must be constrained by measurements, general theoretical considerations, or lattice QCD calculations.
However, these inputs are usually only available within specific kinematic regimes and often need to be extrapolated to the region of interest.
To this end, a parametrisation encapsulating all the relevant features of the FF is required.

In this work, we focus on form factors that describe the transition between two hadrons or the creation of two hadrons from the QCD vacuum.
Such FFs only depend on one dynamical variable, the invariant mass of the two-hadron system $s$.
The most prominent example of such a FF is the pion vector form factor, which enters in a range of physical processes, chief among them pion pair-production in electron--positron collisions, decays of the tau-lepton, and electron--pion scattering.
The first two processes probe the timelike region of the FF, while the latter involves spacelike momentum transfer.
The situation is even more complex for FFs of two-hadron states with different masses, where a third region describing the kinematics of the decay of the heavier to the lighter hadron appears.

More than six decades ago, Meiman~\cite{Meiman:1963kkc} and Okubo~\cite{Okubo:1971jf,Okubo:1971my} introduced a similar FF parametrisation to solve an optimisation problem.
This parametrisation has then been proven to be applicable to systematically parametrise the FFs discussed in this work, in particular for applications in the spacelike or the decay region:
conformally map $s$, such that the entire first Riemann sheet of the FF is contained inside the unit disc, while the branch cut is mapped onto the circle.
FFs can then be expanded in a power series in this new variable.
While the original concept has been improved many times over the years~\cite{Boyd:1997kz,Buck:1998kp,Caprini:1999ws,Bourrely:2008za,Caprini:2017ins,Kirk:2024oyl,Gopal:2024mgb}, the conformal map at its heart has remained the same.

Here, we introduce a set of improved conformal maps that take into account more of the analytic features of the FFs, leading to significantly improved convergence in the timelike production region, as well as allowing for the precise determination of resonance poles and residues, or scattering phases.

The remainder of this paper is structured as follows.
First, we discuss the analytic structure of FFs in \cref{sec::structure}, followed by the introduction of an improved version of the single-threshold map in \cref{sec:conformal}.
In \cref{sec:foursheet}, we present a novel conformal map that takes into account two thresholds.
We conclude in \cref{sec:conclusions}.

\section{Analytic structure of form factors}
\label{sec::structure}

Two-hadron form factors\footnote{%
Here and in the following, the discussion is always meant to be restricted to hadrons stable in QCD.
The analytic structure of the form factors becomes significantly more involved if one of the hadrons is a resonance; cf., e.g., Refs.~\cite{Schneider:2012ez,Ananthanarayan:2014pta} for discussions of the analytic properties of the $\omega\pi$ transition form factor.
} are analytic functions of the complex $s$-plane and satisfy the Schwarz reflection property:
\begin{align}
    F\bigl(s^\ast\bigr) = F^\ast(s)\,.
\end{align}
Consequently, they are real below the first threshold and have a branch cut above.
This threshold corresponds to the energy of the lowest on-shell two-particle state that satisfies the form factor quantum numbers, e.g., $4 M_\pi^2$ for the pion scalar and vector form factors.
The start of the branch cut does not need to coincide with the pair production threshold;
cf., e.g., the case of the vector isovector kaon form factor, for which the first threshold is $4 M_\pi^2$ instead of $4 M_K^2$~\cite{Blatnik:1978wj,Stamen:2022uqh}.
In addition, some form factors possess poles below the branch cut, corresponding to bound states, for example, the $B^\ast$ in the $B\pi$ vector form factor.

Above the first threshold and below the first inelastic threshold, Watson's theorem~\cite{Watson:1952ji} dictates that the phase of the form factor is equal to that of the corresponding partial-wave amplitude.
The threshold behaviour is given by
\begin{align}
    \mathrm{Im}\,F(s) = \mathcal{O}\left((s - s_+)^{(2l+1)/2}\right)\,,
\end{align}
where $l$ denotes the partial wave of the two-hadron system.
Finally, at large $s$, form factors need to drop off sufficiently fast, typically as $1/s$~\cite{Chernyak:1977as,Farrar:1979aw,Efremov:1979qk,Lepage:1979zb,Lepage:1980fj}. 

\begin{figure}[t]
    \centering
    \includegraphics[width=0.75\textwidth]{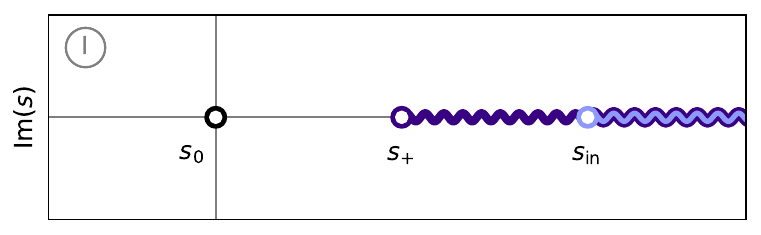}
    \hspace{-7cm}
    \includegraphics[width=0.75\textwidth]{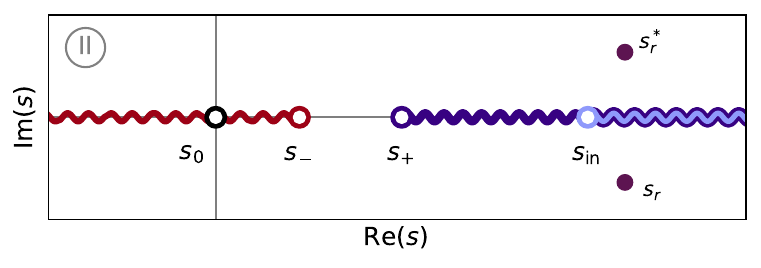}
    \caption{\label{fig:q2plane}
        Analytic structure of a form factor on the first (I) and second (II) Riemann sheets.
        The singularities include a left-hand cut (red line), a unitarity cut opening at $s_+$ (dark blue line) with higher-energy branch cuts (light blue line) opening at the inelastic threshold $s_\text{in}$, as well as resonance poles $s_r$ (purple dots).
    }
\end{figure}

The form factors can be continued onto their second Riemann sheet\footnote{%
In the following, the Riemann sheets will be denoted with numbers according to the branch cut they share.
The first branch cut separates the Riemann sheets I and II, which are further split by the second branch cut into 11, 12, 21, and 22.
}
by identifying
\begin{align}
    F_{\mathrm{II}}(s - i\epsilon) \equiv F_{\mathrm{I}}(s+i\epsilon)
\end{align}
between the first threshold and the inelastic threshold.
This leads to a simple relation between the two Riemann sheets~\cite{Caprini:2017ins}
\begin{align} \label{eq:FF_secondsheet}
    F_{\mathrm{II}}(s) = \frac{F_{\mathrm{I}}(s)}{1 + 2i\,\sigma_\pi(s)\,t(s)} = \frac{F_{\mathrm{I}}(s)}{S(s)}\,,
\end{align}
where $\sigma_\pi(s) = \sqrt{1-4M_\pi^2/s}$ is the appropriate phase-space factor, $t$ the partial-wave amplitude, and $S$ the partial-wave-projected scattering matrix element.
As a consequence, the analytic structure on the second sheet is richer than on the first.
In addition to the regular (right-hand) branch cut (RHC), the appearance of the partial-wave projected scattering amplitude introduces a left-hand cut (LHC) on the second sheet, running from $s_-$ to $-\infty$.\footnote{%
Note that behind the left-hand cut lies another Riemann sheet, which has a new right-hand cut.
This process continues indefinitely, resulting in an infinite sequence of Riemann sheets.
Details can be found in \cref{app:riemann_sheets}.
}
Furthermore, resonances and virtual states correspond to poles on the second sheet of the form factors or, equivalently, to zeros of the $S$-matrix on the first sheet.
While resonance poles appear in complex conjugate pairs, virtual states are situated on the real axis below the RHC, analogous to bound states on the first sheet.
The analytic structure of the form factor on its second sheet is depicted in \cref{fig:q2plane}. 

\subsection[The \texorpdfstring{$z$}{z}-expansion]{The $\boldsymbol z$-expansion}\label{sec:z_expansion}

\begin{figure}[t]
    \centering
    \includegraphics[width=0.5\textwidth]{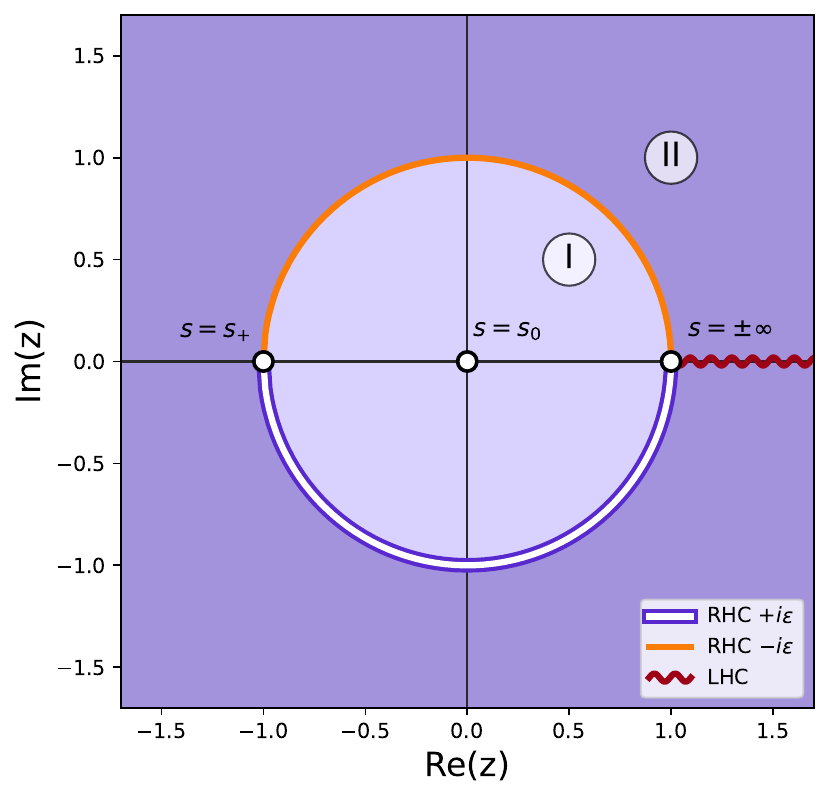}
    \caption{The complex $z$-plane.
        The unfolded right-hand cut (white and orange) separates the physical Riemann sheet (I) from the unphysical Riemann sheet (II).
        For $s_0 = s_-$, the left-hand cut of the second sheet (red) extends from $z = 1$ to $\infty$.
    }
    \label{fig:zmap}
\end{figure}

Simple parametrisations of form factors can be derived by a conformal map from $s$ to $z$ given by
\begin{align}\label{eq:z_map}
    z(s) \equiv z(s;s_+,s_0) = \frac{\sqrt{s_+-s}-\sqrt{s_+-s_0}}{\sqrt{s_+-s}+\sqrt{s_+-s_0}}\,.
\end{align}
Here, $s_+$ is the position of the first threshold, and $s_0 < s_+$ determines the value of $s$ mapped to $z = 0$.
This transformation maps $s_+$ to $z = -1$ and all real values of $s<s_+$ to the real interval $-1 < z < 1$.
The branch cut is mapped to the unit circle and consequently the entire first Riemann sheet is situated within the unit disc, while the second sheet fills the remaining complex plane. The latter can be obtained from
\begin{align}
    z_\mathrm{II}(s) \equiv 1/z(s;s_+,s_0) = \frac{\sqrt{s_+-s}+\sqrt{s_+-s_0}}{\sqrt{s_+-s}-\sqrt{s_+-s_0}}\,.
\end{align}
Both maps can be inverted via
\begin{align}
    s(z;s_+,s_0) = \frac{s_0 (1 + z)^2 - 4 s_+ z}{(1-z)^2}\,.
\end{align}
The poles on the second Riemann sheet lie outside of the unit disc, and their distance from it is determined by their width.
In general, the left-hand cut extends from the negative real value of $z_\mathrm{II}(s_-)$ to $-\infty$ and then back from $\infty$ to $1$, where the points $s = \pm \infty$ of both Riemann sheets are mapped to.
If $s_0 = s_-$, the situation simplifies and the left-hand cut runs from $z_\mathrm{II}(s_-)=\infty$ to $1$.
These properties are summarised in \cref{fig:zmap}.

The variable $z$ is convenient to parametrise form factors if the goal is to describe the region $s < s_+$.
In this case, they can be expressed as
\begin{align}\label{eq:SE}
    F(s) = \frac{1}{\phi_F(z) B_F(z)}\sum_i a_i z^i\,,
\end{align}
where $B_F$ is a Blaschke product with zeros at the positions of the subthreshold bound states and $\phi_F$ is an outer function, i.e., an analytic function defined in terms of its modulus on the unit circle that does not vanish within the disc.\footnote{For explicit definitions, see standard textbooks such as Ref.~\cite{rudin}.} The outer functions are introduced to render the unitarity bounds, which stem from the optical theorem, diagonal on the expansion coefficients $a_i$:
\begin{align}
    \sum_i |a_i|^2 < 1\,.
\end{align}
This series expansion (SE) parametrisation, often referred to as the Boyd--Grinstein--Lebed (BGL) parametrisation~\cite{Boyd:1997kz}, and its simple unitarity bound are powerful tools for describing form factors relevant to semileptonic $B$-meson decays.

Many variations of the BGL parametrisation exist, optimised to different applications.
\begin{itemize}
    \item Setting $\phi_F = 1$ and replacing $B_F$ by a simple pole-factor in $s$, one obtains a simplified series expansion (SSE)~\cite{Bourrely:2008za} (sometimes referred to as Bharucha--Straub--Zwicky (BSZ) parametrisation in reference to the global analysis in Ref.~\cite{Bharucha:2015bzk}).
    While simpler than the SE one, this parametrisation loses the diagonal structure of the unitarity bounds.
    \item The SSE parametrisation can be improved to implement the correct threshold behaviour.
    This has been first derived for the case of $l=1$ form factors, such as the $B\pi$ vector form factor in Ref.~\cite{Bourrely:2008za} and is referred to as the Bourrely--Caprini--Lellouch (BCL) parametrisation.
    \item An additional weight function was introduced to enforce the correct drop-off behaviour for $s \rightarrow \infty$ and at $s = s_+$ for a study of the electromagnetic pion form factor by Buck and Lebed~\cite{Buck:1998kp}.
    \item The impact on the unitarity bounds of poles above threshold was studied in Ref.~\cite{Caprini:2017ins}.
    \item The introduction of explicit poles outside of the unit disc, addressing the convergence issue observed by Buck and Lebed, was studied in Ref.~\cite{Kirk:2024oyl}.
    This approach allowed for a purely data-driven determination of the pole location of the $\rho$-resonance and its residue.
    \item Not allowing for complex zeros in the unit disc improves the convergence in case of the pion vector form factor~\cite{Leutwyler:2002hm,Ananthanarayan:2011xt,RuizArriola:2024gwb,Leplumey:2025kvv}.
\end{itemize}
All of them have in common that only the first Riemann sheet is contained in the unit disc, and thus they are not suitable for describing the form factors for $s > s_+$.

\section{Improved conformal expansions}\label{sec:conformal}

To describe the form factors beyond the first inelastic threshold, one needs to modify the conformal map suggested in \cref{eq:z_map}.
We first present an improvement of the usual SE to account for the LHC and to ensure convergence of the series in the region of interest.
To test our parametrisation, we fit it to pseudo-data for the scalar isoscalar and the vector isovector pion form factors that we sample via the Omnès representation~\cite{Omnes:1958hv} with phase input from the (modified) inverse amplitude method.

\subsection[The left-hand cut and the \texorpdfstring{$\zeta$}{ζ}-map]{The left-hand cut and the $\boldsymbol\zeta$-map} \label{sec:zeta_map}
The radius of convergence of the usual SE in \cref{eq:SE} is limited to 1 both by higher-energy branch cuts that are mapped to symmetric arcs of the unit circle and by the LHC of the second Riemann sheet, which is usually mapped to $z \in [1, \infty]$.
In particular, the usual SE is not expected to be valid in any region of the second sheet.

In practice, the coefficients of the SE are fitted to experimental or theoretical inputs, and the domain of validity of the parametrisation is non-trivially deformed by numerical instabilities and oscillations due to Runge's phenomenon.
As found in Ref.~\cite{Buck:1998kp}, the presence of resonance poles on the second sheet, close to the unit circle, can, for example, drastically affect the region of convergence of the SE.
Although explicitly parametrising these poles significantly improves the description of the data~\cite{Kirk:2024oyl}, relating the pole parameters to the physical parameters of the $T$-matrix requires the SE to be valid at the pole position, which, as described above, is not ensured by the usual $z$-map.

As we will show below, the parameters of the poles that are close to the unit circle, such as the $\rho(770)$-resonance of the vector pion form factor, can still be correctly described by their imprint on the first sheet.
Poles that are further away from the region of convergence, however, suffer from blind directions in the fit, which make the extraction of their parameters from data more challenging.

\begin{figure}[t]
    \centering
    \begin{subfigure}{0.48\textwidth}
        \includegraphics[width=\textwidth]{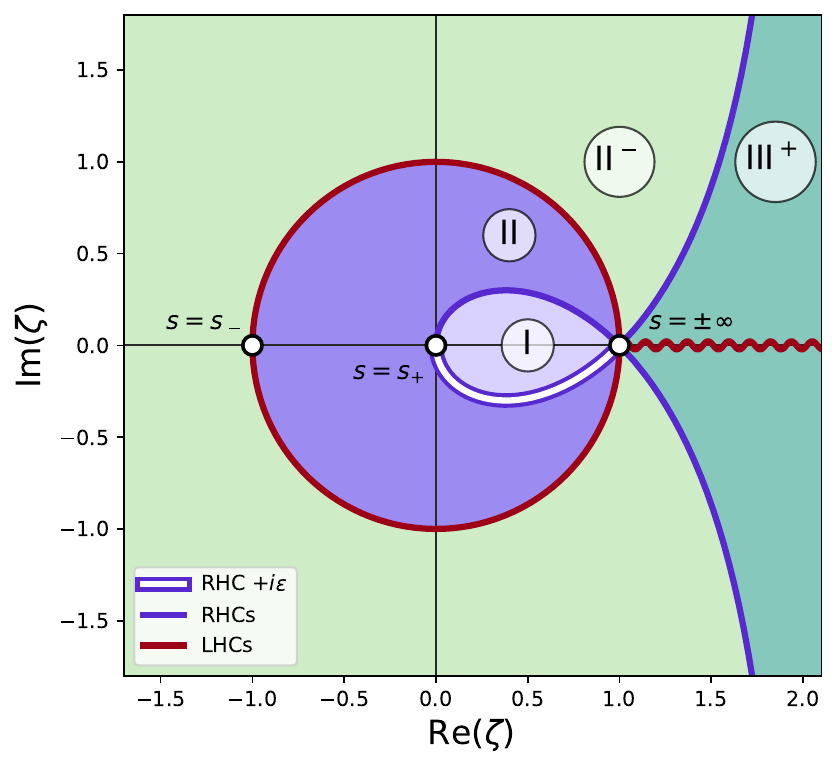}
    \end{subfigure}
    \begin{subfigure}{0.48\textwidth}
        \includegraphics[width=\textwidth]{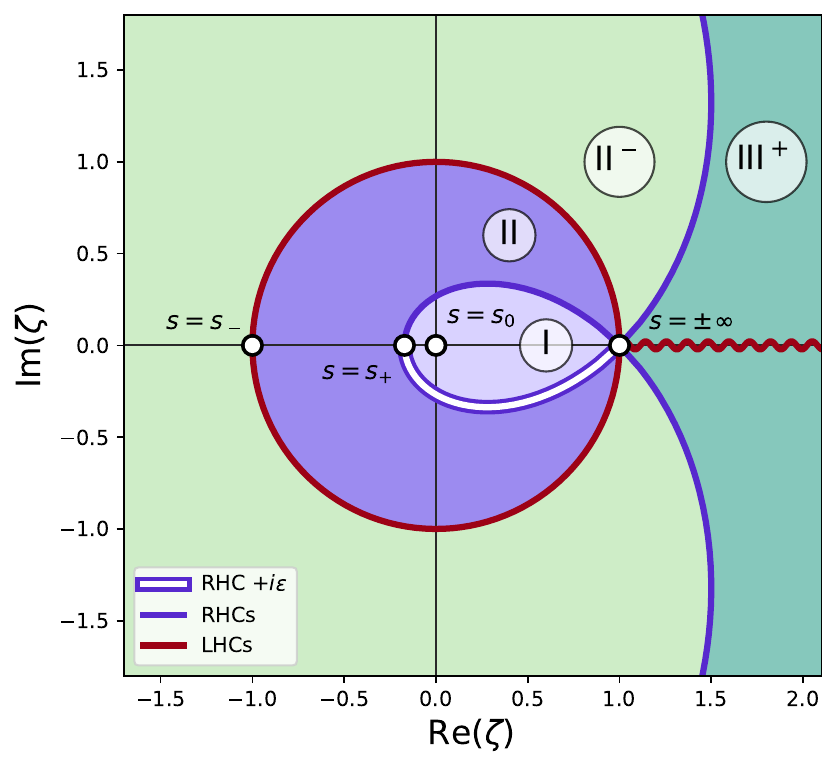}
    \end{subfigure}
    \caption{\label{fig:zeta_map}
        The complex $\zeta$-plane for $z_0 = -1$ (left) and $z_0 = 0$ (right).
        Four sheets are visible on the plane, sheets I and $\text{II} \equiv \text{II}^+$ are separated by the RHC, $\text{II}^+$ and $\text{II}^-$ by the LHC, and $\text{II}^-$ and $\text{III}^+$ by a RHC.
        Another LHC still spans $\zeta = 1$ to $\infty$, see \cref{app:riemann_sheets} for details. 
        Higher-energy branch cuts due to other channels are not represented.
        The physical timelike region where the scattering data lies is highlighted in white.
    }
\end{figure}

A more promising approach consists of modifying the conformal map in \cref{eq:z_map} to account for the LHC explicitly.
By doing so, we ensure that singularities only stem from parametrisable poles and higher-order branch cuts.
This can be achieved by applying the same idea as for the pair-production branch cut in $s$.
Concretely, we define the following conformal map
\begin{equation} \label{eq:cmap_definition}
    \omega(x,x_0) = \frac{\sqrt{1-x_0}-\sqrt{1-x}}{\sqrt{1-x_0}+\sqrt{1-x}}\,,
    \qquad
    \omega^{-1}(y,x_0) = \frac{x_0 (1 - y)^2 + 4 y}{(1+y)^2}\,.
\end{equation}
With this definition, the $z$-map of \cref{eq:z_map} can be written as
\begin{align}
    z(s;s_+,s_0) = -\omega\biggl(\frac{s}{s_+},\frac{s_0}{s_+}\biggr)\,.
\end{align}
Applying the $\omega$-map allows us to unfold the left-hand cut and the second-sheet right-hand cuts successively.
We call this new map $\zeta$ and define it as
\begin{align} \label{eq:zeta_map_definition}
    \zeta(s)\equiv\zeta(s;s_+,z_0) = \omega\bigl(z(s;s_+,s_0=s_-),z_0\bigr) = \omega\biggl(-\omega\biggl(\frac{s}{s_+},\frac{s_-}{s_+}\biggr),z_0\biggr)\,,
\end{align}
where $s_0 = s_-$ is chosen to map the LHC on the $z \in [1, \infty]$ range and $z_0$ is a free parameter that satisfies $\omega\bigl(z_0, z_0\bigr) = 0$.
The analytic structure of the form factor in the $\zeta$-plane is sketched in \cref{fig:zeta_map}.
In this variable, the point $s=\infty$ corresponds to $\zeta=1$, the first (I) and second (II) Riemann sheets are mapped inside the unit disc, and the left-hand cut spans the unit circle.
Two Riemann sheets are accessible outside the disc.
Following the naming scheme of Ref.~\cite{Jing:2025qmi}, they are labelled $\text{II}^-$ and $\text{III}^+$ and are separated by a right-hand cut, already opened by the $\zeta$-map.
More information on how these sheets are connected is provided in \cref{app:riemann_sheets} and in Ref.~\cite{Jing:2025qmi}.
Higher-energy branch cuts due to heavier final states (e.g., $K\bar{K}$ or $\pi\omega$ in the case of the pion form factors) are still present between the first and the second sheets.
Finally, as detailed in \cref{app:riemann_sheets}, a new left-hand cut extends from $\zeta=1$ to $\infty$.

The second Riemann sheet in $s$ can be reached via
\begin{align}
    \zeta_\text{II}(s) \equiv \zeta_\text{II}(s; s_+, z_0) = \omega\bigl(z_\text{II}(s; s_+, s_-), z_0\bigr)\,,
\end{align}
while the ones behind the left-hand cut are given by
\begin{align}
    \zeta_{\text{II}^-}(s)  = \frac{1}{\zeta_\text{II}(s)}\,, \qquad
    \zeta_{\text{III}^+}(s) = \frac{1}{\zeta(s)}\,.
\end{align}
As for the $z$-map, all four sheets also share a common inverse, which is defined by
\begin{align}
    s(\zeta)=s_+ \, \omega^{-1}\biggl(-\omega^{-1}(\zeta,z_0),\frac{s_-}{s_+}\biggr)\,.
\end{align}

With this map, we can write the form factors using a simplified $\zeta$ series expansion ($\zeta$ SSE)\footnote{%
Although outer functions can also be used in the parametrisation, making the unitarity bound manifestly diagonal is made challenging by the non-trivial integration path in the $\zeta$-plane.
We follow, therefore, the same argument as in Ref.~\cite{Bourrely:2008za} and only keep the part of the SE that reproduces the analytic behaviour of the series.
}
\begin{align}\label{eq:zeta_SSE}
    F(s) = \mathcal{P}(\zeta) \sum_i a_i \zeta^i ~ \Big|_{\zeta = \zeta(s)} \,,
\end{align}
where $\mathcal{P}(\zeta)$ implements the poles of $F$.
Resonances give conjugate poles on the unphysical sheets, while bound states and virtual states are implemented as simple poles on the real axis of the physical and the unphysical sheets, respectively,
\begin{equation}
    \mathcal{P}(\zeta) = \prod_{k\in\{\text{bound states}\}} \frac{1}{\zeta - \zeta_k} ~ \prod_{l\in\{\text{virtual states}\}} \frac{1}{\zeta - \zeta_l} ~ \prod_{m\in\{\text{resonances}\}} \frac{1}{(\zeta - \zeta_m)(\zeta - \zeta^*_m)} \,. 
\end{equation}
Although the remaining LHC only starts at $\zeta = 1$, the region of convergence of this series is limited by higher-energy branch cuts to $|\zeta(s)| < \zeta(s_\mathrm{in})$, where $s_\mathrm{in}$ is the squared invariant mass of the first inelastic threshold.
This means that, contrary to the usual SE, the $\zeta$-expansion \cref{eq:zeta_SSE} is not valid on the full physical sheet.
On the other hand, the timelike region we are trying to describe now lies within the theoretical region of convergence of the series, rather than at its boundary.
The resonance poles are now also inside the unit disc and can be implemented using a product of factors, in complete analogy to the subthreshold poles in \cref{eq:SE}.
Since the lowest-lying resonances are now within the region of convergence of the series, their parameters can be fitted and extracted from the data.
Finally, at large $s$ (i.e., $\zeta\to1$), we have the following relation 
\begin{align}
    \left(\zeta(s)-1\right)^4 \underset{s\to\infty}{\sim} \frac{1}{s}\,, 
\end{align}
which can be relevant to fix the high-energy behaviour of form factors.

We conclude this section with the remark that, in principle, one could repeat the procedure and keep applying the $\omega$-map to open up the new left-hand cuts on the sheets outside the unit circle, as described in more detail in \cref{app:zeta}.
This might result in a more accurate description of the second-sheet left-hand cut, as it is shifted further into the radius of convergence, provided that higher-energy thresholds can be described accurately.

\begin{table}[t]
    \centering
    \begin{tabular}{cccc}
        \toprule
        resonance $r$ & $\sqrt{s}_r\,[\mathrm{GeV}]$ & $z_r$ & $\zeta_r$ \\
        \midrule
        $f_0(500)$ & $0.4515-0.2266i$ & $0.956-1.459i$ & $-0.110-0.041i$ \\
        $\rho(770)$ & $0.7614-0.0755i$ & $0.796-0.730i$ & $\phantom{-}0.077-0.335i$ \\
        \bottomrule
    \end{tabular}
    \caption{Comparison between the pole positions of the $f_0(500)$ and $\rho(770)$ resonances as produced by the IAM (mIAM) in the different conformal planes, $z_r=z_\mathrm{II}(s_r;s_+=4M_\pi^2,s_0=0)$ and $\zeta_r=\zeta_\mathrm{II}(s_r;s_+=4M_\pi^2,z_0=0)$.}
    \label{tab:polepositions_SC}
\end{table}

\subsection{Application to single-channel pion form factors}\label{sec:application_SingleChannel}

As a concrete example, we test the effectiveness of the $\zeta$-map on the scalar isoscalar and vector isovector pion form factors.
We generate pseudo-data using the next-to-leading-order (NLO) inverse amplitude method (IAM)~\cite{Niehus:2020gmf} for the phases $\delta_l^I(s)$ and an Omnès representation~\cite{Omnes:1958hv}
\begin{equation}
    \Omega_l^I(s) = \exp\Biggl( \frac{s}{\pi} \int_{4M_\pi^2}^{\infty} \frac{\d s^\prime}{s^\prime} \frac{\delta_l^I(s^\prime)}{s^\prime-s} \Biggr)\,.
\end{equation}
Our theoretical model only includes the $\pi\pi$ channel and the lowest-lying resonances, namely the $f_0(500)$ and the $\rho(770)$, respectively.
The advantage of using generated data is that the resonance parameters can be extracted exactly from the IAM and compared with the values obtained from the fits in $\zeta$ and $z$.
We provide the position of the poles in the variables $z$ and $\zeta$ in \cref{tab:polepositions_SC}.
For the scalar isoscalar pion form factor, we employ a modified version of the IAM (mIAM) that can accurately reproduce the location of the Adler zero~\cite{Pelaez:2010fj, GomezNicola:2007qj}.
More details on the two models are provided in \cref{app:th_models}.
A complete analysis of the data, including uncertainties, is kept for future work.

Our test parametrisations are truncated versions of \cref{eq:zeta_SSE}
\begin{align}
    F_\mathrm{fit}^{(n)}(s) = \frac{1}{(\zeta - \zeta_p)(\zeta - \zeta_p^*)} \sum_{k=0}^n a_k \zeta^k ~ \Big|_{\zeta = \zeta(s)} \,,
\end{align}
where $\zeta_p\equiv\zeta_\mathrm{II}(s_p)$ is the position of the $f_0(500)$ or $\rho(770)$ pole in the $\zeta$-plane.
In this analysis, we impose $s_+ = 4M_\pi^2$, $s_0 = s_-$, and $z_0 = 0$, while fitting the complex pole location $\sqrt{s_p}$ and the coefficients $a_k$.
A similar parametrisation, obtained by replacing $\zeta$ with $z$, is also used for comparison purposes.\footnote{%
This parametrisation slightly differs from that of Ref.~\cite{Kirk:2024oyl}, as we do not impose high-energy or threshold behaviour on the series.
However, as in Ref.~\cite{Kirk:2024oyl}, we use $s_0 = -1\,\mathrm{GeV}^2$.}

The pseudo-data consist of 50 data points sampled in the energy interval $s \in [0, 1]\,\mathrm{GeV}^2$ for the scalar form factor and in  $s \in [0, 1.2]\,\mathrm{GeV}^2$ for the vector one.
We fit this data with the $z$ and $\zeta$ SSE using different values of the truncation order, $n = 3, 4$, and $5$.
From the best-fit points, we extract the pole positions of the $\sigma \equiv f_0(500)$ and $\rho \equiv \rho(770)$ resonances, as well as their couplings to pions, photons, or light quarks (as detailed in \cref{app:pole_parameters}).
We test two scenarios.
In the first case, we fit our parametrisation to the real and imaginary parts of the theoretical model to test whether the latter can be accurately reproduced.
In the second case, we only fit our parametrisation to the modulus of the theoretical model, to mimic an analysis on experimental data.
Since both scenarios yield very similar results, both in terms of pole positions and expansion coefficients, we will only discuss the results of the first scenario.
The results are displayed in \cref{tab:Fits_SFF_SingleChannel,tab:Fits_VFF_SingleChannel}, where we also provide an indicative $\chi^2$ obtained by assuming the input points to be uncorrelated and with an uncertainty of 1.%
\begin{table}[t]
    \centering
    \resizebox{\columnwidth}{!}{%
    \begin{tabular}{cccccccc}
        \toprule
        & $n$ & $\chi^2$ $[10^{-6}]$ & $\sqrt{s_\sigma}$ [GeV] & $|g_{\sigma\pi\pi}|$ [GeV] & $\arg(g_{\sigma\pi\pi})$ [$^\circ$] &  $|g_{\sigma q\bar q}|$ [GeV] & $\arg(g_{\sigma q\bar q})$ [$^\circ$] \\
        \midrule
        exact & -- & -- & $0.4515-0.2266i$ & $2.895$ & $-66.07$ & $0.100$ & $-12.76$ \\
        \midrule
            & 3 & $67.8$ & $0.4511-0.2216i$ & $2.814$ & $-66.59$ & $0.098$ & $-13.34$ \\
            $z$ & 4 & $1.45$ & $0.4493-0.2267i$ & $2.912$ & $-67.01$ & $0.101$ & $-13.60$ \\
            & 5 & $1.02$ & $0.4498-0.2275i$ & $2.926$ & $-66.79$ & $0.101$ & $-13.37$ \\
        \midrule
            & 3 & $1.43$ & $0.4514-0.2260i$ & $2.886$ & $-66.16$ & $0.100$ & $-12.86$ \\
            $\zeta$ & 4 & $0.018$ & $0.4514-0.2266i$ & $2.898$ & $-66.13$ & $0.100$ & $-12.82$ \\
            & 5 & $0.013$ & $0.4514-0.2267i$ & $2.898$ & $-66.10$ & $0.100$ & $-12.79$ \\
        \bottomrule
    \end{tabular}
    }
    \caption{\label{tab:Fits_SFF_SingleChannel}
        Fit summary for the $z$ and $\zeta$ SSE to the single-channel $I=0$, $l=0$ form factor compared to the exact results from the mIAM.
        The quoted $\chi^2$ is computed by assuming uncorrelated uncertainties of 1 for all 50 input data points.
        The last five columns correspond to the $f_0(500)$ pole parameters extracted from the different models.
    }
\end{table}
\begin{table}[t]
    \centering
    \resizebox{\columnwidth}{!}{%
    \begin{tabular}{cccccccc}
        \toprule
        & $n$ & $\chi^2$ $[10^{-6}]$ & $\sqrt{s_\rho}$ [GeV] & $|g_{\rho\pi\pi}|$ & $\arg(g_{\rho\pi\pi})$ [$^\circ$] &  $|g_{\rho \gamma}|$ & $\arg(g_{\rho \gamma})$ [$^\circ$] \\
        \midrule
        exact & -- & -- & $0.76143-0.07548i$ & $6.0389$ & $-6.951$ & $5.4979$ & $-3.828$ \\
        \midrule
                & 3 & $698$ & $0.76123-0.07551i$ & $6.0433$ & $-7.092$ & $5.4927$ & $-3.767$ \\
            $z$ & 4 & $144$  & $0.76142-0.07558i$ & $6.0463$ & $-6.947$ & $5.4903$ & $-3.823$ \\
                & 5 & $18.0$  & $0.76147-0.07548i$ & $6.0388$ & $-6.919$ & $5.4952$ & $-3.840$ \\
        \midrule
                    & 3 & $215$   & $0.76132-0.07548i$ & $6.0405$ & $-7.028$ & $5.4944$ & $-3.794$ \\
            $\zeta$ & 4 & $12.7$  & $0.76142-0.07551i$ & $6.0409$ & $-6.950$ & $5.4935$ & $-3.822$ \\
                    & 5 & $0.253$ & $0.76143-0.07548i$ & $6.0389$ & $-6.948$ & $5.4948$ & $-3.823$ \\
        \bottomrule
    \end{tabular}
    }
    \caption{\label{tab:Fits_VFF_SingleChannel}
        Fit summary for the $z$ and $\zeta$ SSE to the single-channel $I=1$, $l=1$ form factor compared to the exact results from the IAM.
        The quoted $\chi^2$ is computed by assuming uncorrelated uncertainties of 1 for all 50 input data points.
        The last five columns correspond to the $\rho(770)$ pole parameters extracted from the different models.
    }
\end{table}
For a given truncation order, we find that the $\zeta$ SSE always performs better than the $z$ one, in terms of $\chi^2$, position, and coupling of the resonances.
With $n = 4$, all the resonance parameters are correctly captured by the $\zeta$-expansion, while this is still not the case for the $z$ one with $n = 5$.
This suggests that the convergence of the $\zeta$-expansion is very fast, even when fitting data above the elastic threshold.

For completeness, the modelled form factors and their best $n = 3$ fits are displayed in \cref{fig:Fits_SingleChannel}.
The best-fit parameters are also provided in the supplementary material.
\begin{figure}[t]
    \centering
    \begin{subfigure}{0.5\textwidth}
        \includegraphics[width=\textwidth]{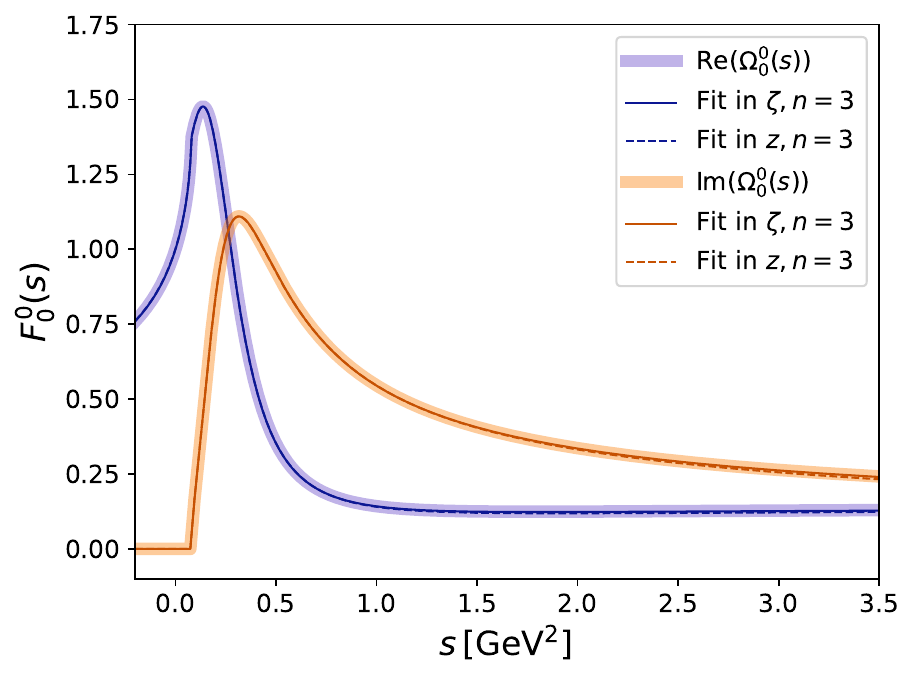}
    \end{subfigure}
    \begin{subfigure}{0.49\textwidth}
        \includegraphics[width=\textwidth]{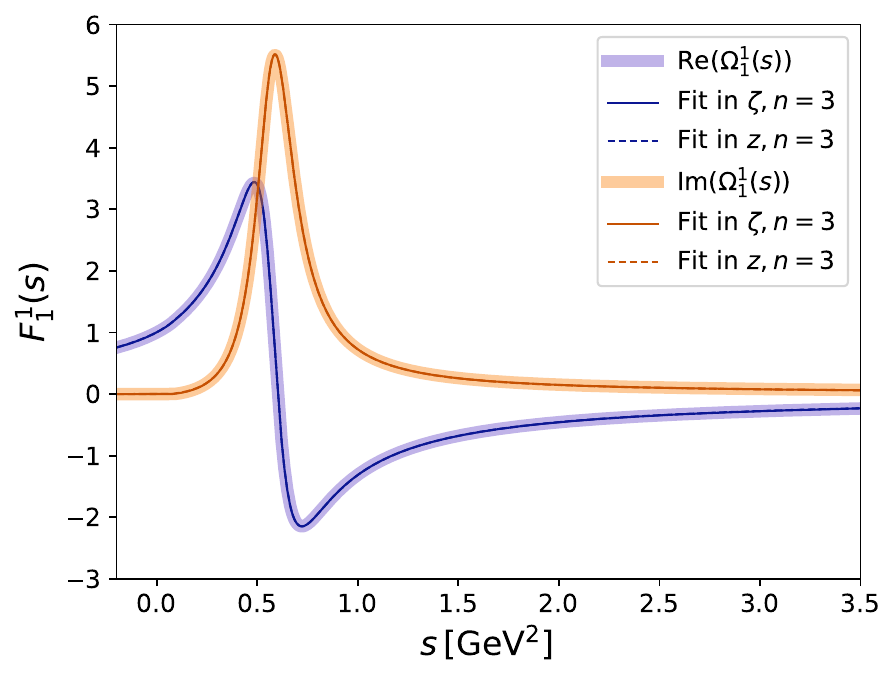}
    \end{subfigure}
    \caption{\label{fig:Fits_SingleChannel}
        Comparison between the modelled form factors $\Omega_l^I$ and the SSE fits to pseudo data.
        The fits are performed in the regions $s\in[0,1]\,\mathrm{GeV}^2$ for the scalar form factor $I=0$, $l=0$ (left) and $s\in[0,1.2]\,\mathrm{GeV}^2$ for the vector form factor $I=1$, $l=1$ (right).
    }
\end{figure}
In the physical region, all the fits visually agree perfectly with the data.
We recall that these tests are intended as a proof of concept to demonstrate the possibilities that this new variable offers.
A more in-depth analysis of the data, including the extraction of the pole positions, is beyond the scope of the present manuscript.

\subsection{Conformal parametrisation for partial-wave amplitudes}
One significant advantage of the $\zeta$-parametrisations over those based on $z$ is that the former's domain of validity extends to part of the second Riemann sheet.
Consequently, we can use the relationship between the form factors and the partial-wave projections of the $S$-matrix elements to parametrise these functions simultaneously.
Rewriting \cref{eq:FF_secondsheet}, we have the following parametrisation for the $\pi\pi$ partial-wave amplitude in terms of the form factor on its first and second sheets,
\begin{equation}\label{eq:def_TfromF}
    t(s) = \frac{1}{2\bar\sigma_\pi(s)} \biggl( 1 - \frac{F_\mathrm{I}(s)}{F_\mathrm{II}(s)} \biggr)\,,
\end{equation}
where $\bar{\sigma}_\pi(s)=\sqrt{4M_\pi^2/s-1}$ is the correct analytic continuation of the usual phase-space factor $\sigma_\pi(s) = \sqrt{1-4M_\pi^2/s}$, related by $\bar\sigma_\pi(s\pm i\epsilon)=\mp i \sigma_\pi(s)$ along the RHC~\cite{Moussallam:2011zg, Hoferichter:2017ftn}.
While access to the second sheet is usually difficult to obtain, it is manifestly built into our parametrisation; simply by replacing $\zeta(s) \mapsto \zeta_\mathrm{II}(s)$ in \cref{eq:zeta_SSE}. The $\zeta$-parametrisation, by construction, fulfils the following desired properties:
\begin{itemize}
    \item The phases of $t(s)$ and $F(s)$ match exactly along the unitarity cut, therefore automatically ensuring unitarity in the form of Watson's theorem~\cite{Watson:1952ji}.
    \item A resonance pole $s_r$ on the second sheet of the form factor $F_\mathrm{II}(s)$ automatically produces a zero for the $\pi\pi\to\pi\pi$ $S$-matrix element $S(s)=1-2\,\bar\sigma_\pi(s)\,t(s)$ and, thus, a pole on the second sheet $S_\mathrm{II}(s)$ at the same position $s_r$.
    \item By construction of the conformal variable $\zeta$ a left-hand cut for $F_\mathrm{II}(s)$ and, thus, also for $t(s)$ is produced.
    This is not the case if $z$ is used instead of $\zeta$ as the conformal variable.
\end{itemize}
On the other hand, since we do not constrain $F_\mathrm{II}(0) = 0$, \cref{eq:def_TfromF} imposes the partial-wave amplitude to vanish at $s = 0$ instead of at its Adler zero.
Yet, the combination of these properties offers the possibility to examine the form factor and the partial-wave scattering amplitude within a single framework.

\begin{figure}[t]
    \centering
    \begin{subfigure}{0.5\textwidth}
        \includegraphics[width=\textwidth]{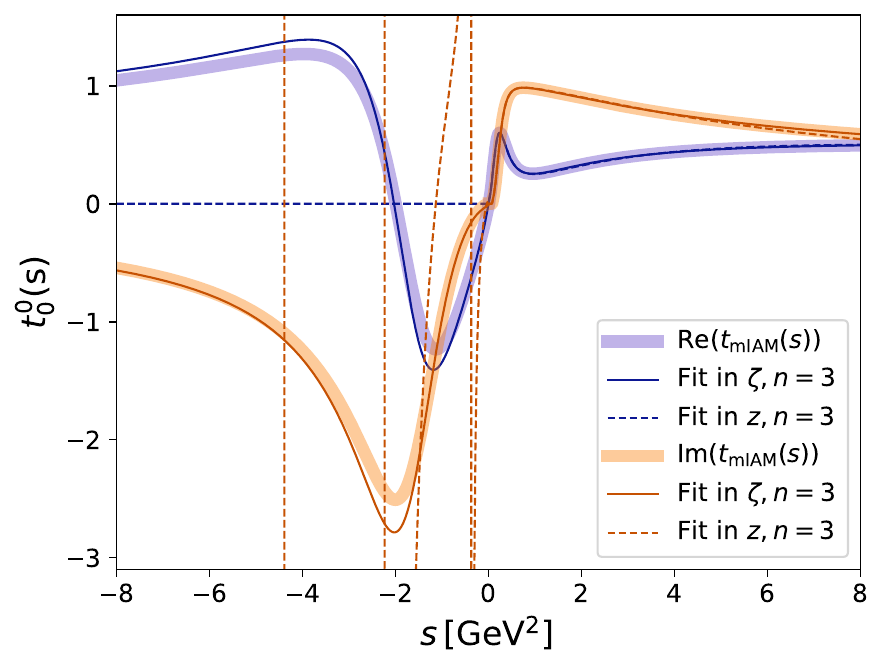}
    \end{subfigure}
    \begin{subfigure}{0.49\textwidth}
        \includegraphics[width=\textwidth]{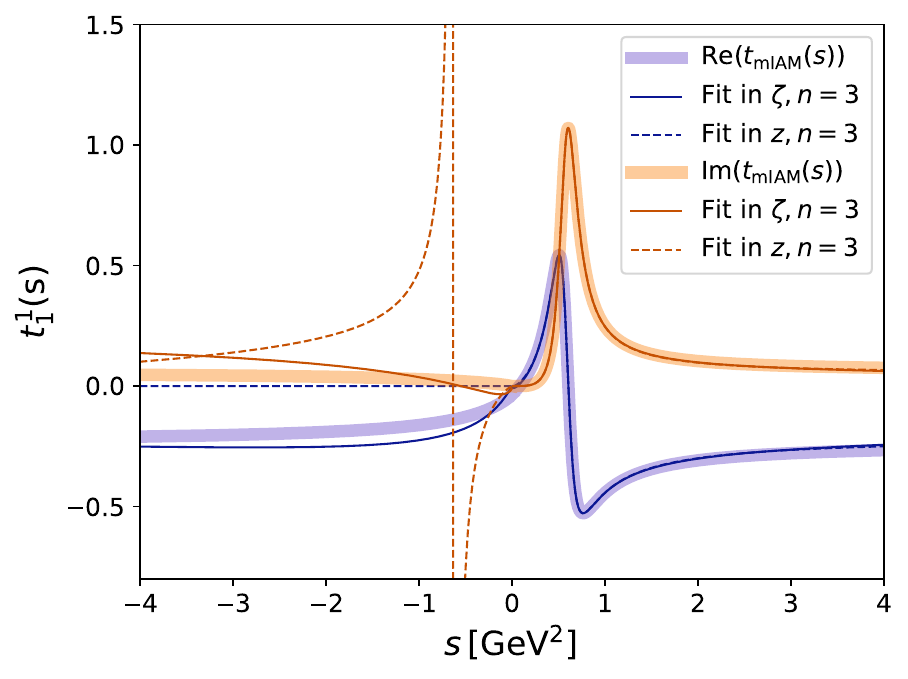}
    \end{subfigure}
    \caption{\label{fig:Fits_SFF_FF_to_t}
        Predictions for the partial-wave amplitudes from fits to the theoretical form factors $\Omega_0^0$ (left) and $\Omega_1^1$ (right).
        The form factors are fitted in the ranges $s \in [0,1]\,\mathrm{GeV}^2$ and $s \in [0,1.2]\,\mathrm{GeV}^2$ respectively.
        Fitting with $n = 3$ truncated $z$ (dashed) and $\zeta$ (plain) SSEs, we find that only the one based on $\zeta$ gives physical results.
    }
\end{figure}

Firstly, one can predict the partial-wave amplitude from a fit to the form factor.
This is particularly challenging, as it requires an accurate description of the form factor on both the first and the second Riemann sheets.
We provide an example of the approach in \cref{fig:Fits_SFF_FF_to_t}, where the $I=0$, $l=0$ and $I=1$, $l=1$ partial waves are predicted from the best $n=3$ fits discussed above.
Clearly, using the $z$-parametrisation leads to completely irrelevant results.
The $\zeta$ SSE performs very well above threshold, as expected from Watson's theorem, but deviates from the model below threshold, especially close to the LHC branch point.

\begin{figure}[t]
    \centering
    \begin{subfigure}{0.5\textwidth}
        \includegraphics[width=\textwidth]{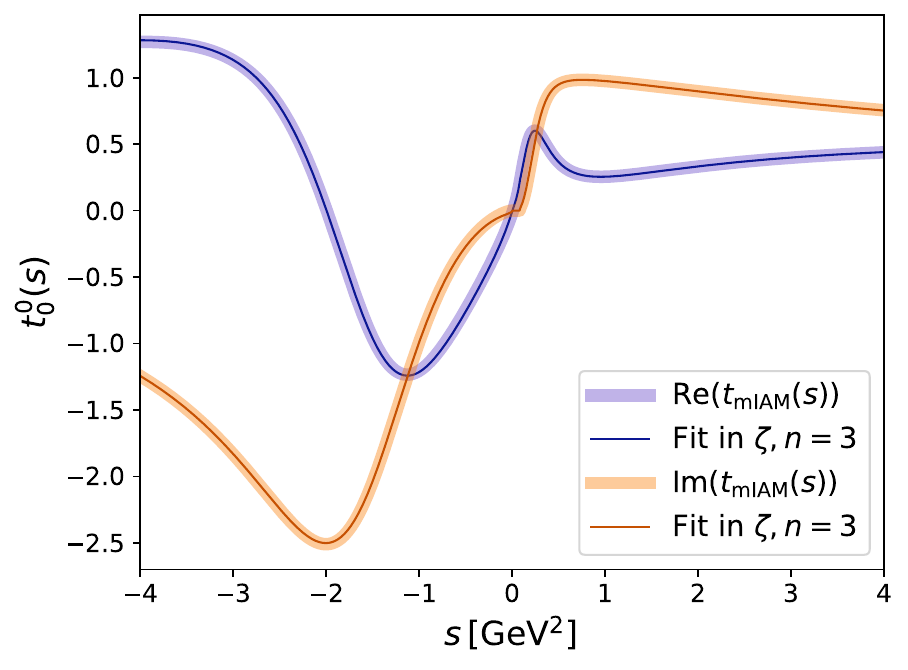}
    \end{subfigure}
    \begin{subfigure}{0.49\textwidth}
        \includegraphics[width=\textwidth]{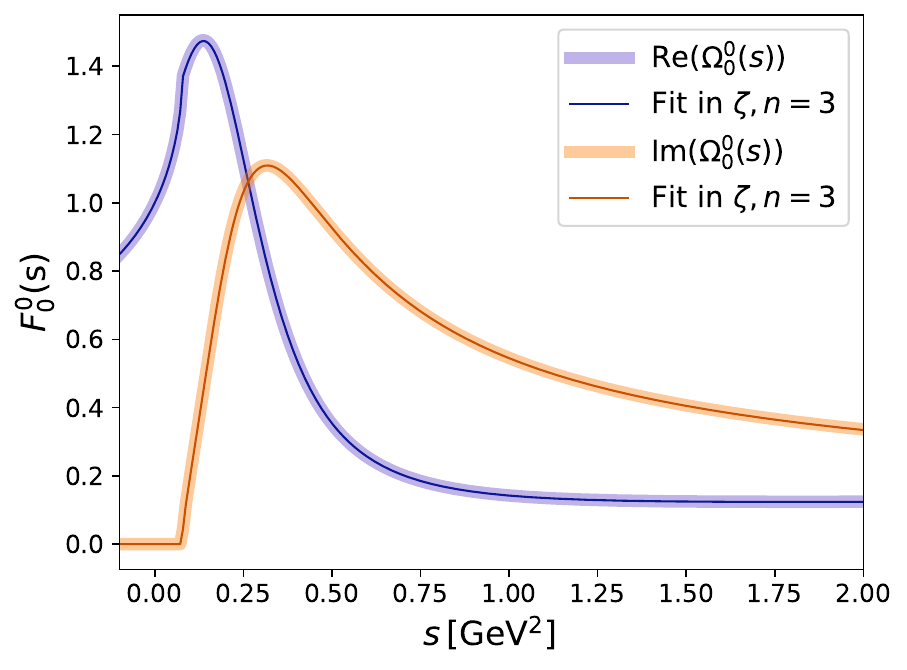}
    \end{subfigure}
    \caption{\label{fig:Fits_SFF_to_T}
        Results of a fit of the partial-wave $t^0_0(s)$ from the mIAM model using the parametrisation from \cref{eq:def_TfromF} with a $n = 3$ truncated $\zeta$ SSE (left).
        The model is fitted in the range $s\in[-4,4]\,\mathrm{GeV}^2$.
        The resulting predicted form factor shows perfect agreement with the theoretical one (right).
    }
\end{figure}

Secondly, the approach can be reversed by predicting the form factor shape from a fit to the scattering data.
This is illustrated in \cref{fig:Fits_SFF_to_T}, where the left plot shows our fit to the $I=0$, $l=0$ scattering data and the right plot shows the corresponding predicted form factors.
The excellent agreement shows the clear advantage of describing the first two sheets using a parametrisation that ensures unitarity.
Similar results were obtained in the vector case, albeit with small shifts at the resonance poles.

Finally, the fit can be performed on both datasets simultaneously, which should, if performed on experimental data, considerably reduce the form factor and partial-wave uncertainties.

\section{Beyond the inelastic threshold} \label{sec:foursheet}
We have shown that, by opening the LHC, the parametrisations based on the $\zeta$-map allow for a much more accurate description of the form factors in the physical region.
In practice, however, their range of validity is limited by the presence of higher-energy branch cuts due to the opening of inelastic thresholds.
For example, we do not expect the parametrisation of the pion form factors presented in the previous section to be valid above $s_\mathrm{in} \approx 1 \,\mathrm{GeV}^2$, where the $K\bar{K}$ and $\pi\omega$ inelasticities become large.

Extending the validity of the parametrisations based on a $z$-map was already suggested in Ref.~\cite{Caprini:1999ws}, where a hybrid approach relying on Omnès functions~\cite{Omnes:1958hv} was suggested.
The $z$-map is performed with the inelastic threshold $s_\text{in}$ instead of the first threshold $s_+$, while an Omnès function describes the branch cut in the elastic regime.
In Ref.~\cite{Herren:2025cwv}, this approach was combined with the correct threshold-scaling at the inelastic threshold following BCL to describe the $\pi\pi$ invariant mass spectrum in $B\rightarrow \pi\pi\ell\nu$ decays.
The main drawback of this approach is that the effect of resonances on the second sheet is encoded in the phase shifts that enter the Omnès functions and cannot be directly extracted from the form factors.

In the following, we present a new strategy that combines the strengths of the $\zeta$-map with the ideas of the hybrid Omnès--$z$-map approach.

\subsection{Conformal 4-sheet map}
\begin{figure}[t]
    \centering
    \includegraphics[width=0.5\textwidth]{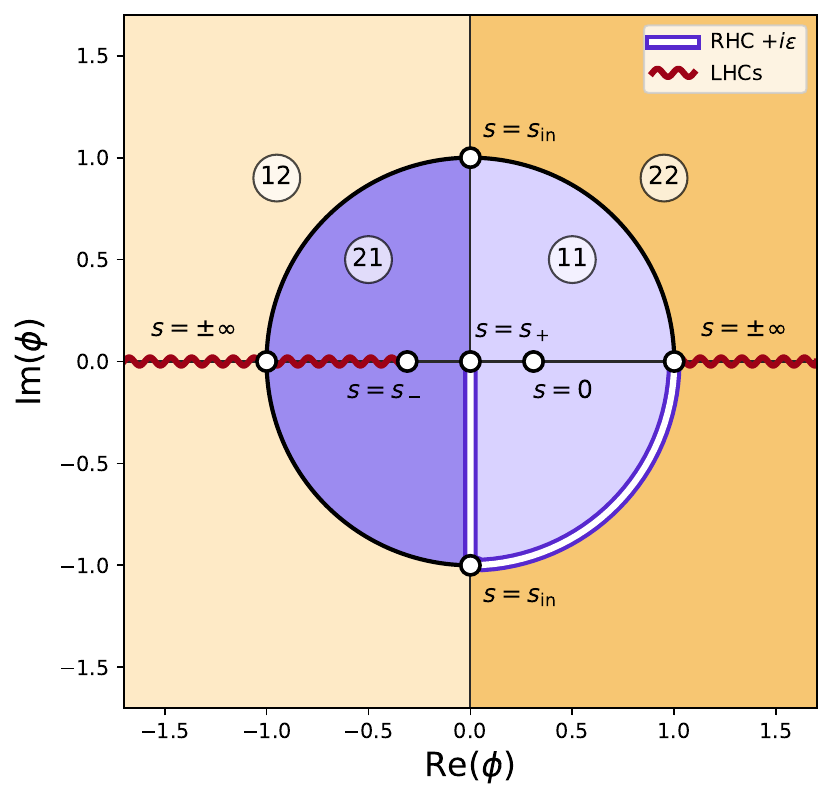}
    \caption{The complex $\phi$-plane.
        The first (11) and second (21) Riemann sheets span the unit disc.
        The unfolded inelastic branch cut is mapped to the disc and separates these two sheets from the further unphysical sheets (12) and (22).
        All the unphysical sheets contain LHCs (in red).
        The physical timelike region where the scattering data lies is highlighted in white.}
    \label{fig:phi_map}
\end{figure}

Following the idea from Ref.~\cite{Caprini:1999ws} and replacing not only $s_+$ from the $z$-map with $s_\text{in}$, but also $s_0\rightarrow s_+$ and $z\rightarrow\phi^2$, leads us to the inverse of a conformal variable $\phi$, given by
\begin{align}
      s(\phi)\equiv s(\phi;s_+,s_\mathrm{in}) = \frac{s_+ (1 + \phi^2)^2 - 4 s_\text{in} \phi^2}{(1-\phi^2)^2}\,,
\end{align}
which now incorporates not one but two thresholds.
The variable $\phi$ therefore opens up two unitarity cuts, and maps the corresponding four Riemann sheets to distinct regions in the complex plane.
For the different Riemann sheets, the $\phi$-map reads
\begin{align}\label{eq:phi_map}
    \phi_{(11)}(s)\equiv\phi_{(11)}(s;s_+,s_\mathrm{in})&=\frac{\sqrt{s_\text{in}-s}-\sqrt{s_\text{in}-s_+}}{\sqrt{s_+-s}}=-\phi_{(21)}(s;s_+,s_\mathrm{in})\equiv-\phi_{(21)}(s)\,, \notag\\
    \phi_{(22)}(s)\equiv\phi_{(22)}(s;s_+,s_\mathrm{in})&=\frac{\sqrt{s_\text{in}-s}+\sqrt{s_\text{in}-s_+}}{\sqrt{s_+-s}}=-\phi_{(12)}(s;s_+,s_\mathrm{in})\equiv-\phi_{(12)}(s)\,,
\end{align}
which can also be obtained by taking the square root of $z$ and using the same associations for the thresholds as above.
A graphic representation of the complex $\phi$-plane can be found in \cref{fig:phi_map}.
With $\phi$, the (11) (also referred to as the physical or first sheet) and (21) sheets are mapped to the inside of the unit circle, while the (12) and (22) sheets are mapped to the outside.
This is the key difference to a similar map introduced in Refs.~\cite{Kato:1965iee,Badalian:1981xj}, where the (11) sheet lies outside of the unit circle and instead the (12) sheet lies inside, thus prohibiting a useful power series expansion.
The first and second threshold, $s_+$ and $s_\text{in}$, respectively correspond to $\phi = 0$ and $\phi = \pm i$.
Similarly, the point $s = 0$ is mapped to the real axis between 0 and 1 or $-1$ and 0 depending on whether the (11) or (21) sheet is considered, and the points $s = \pm\infty$ meet at $\phi = \pm 1$.
The physical axis, approaching the cuts from above, follows the white line pictured specifically in \cref{fig:phi_map}.
In this figure, the connection between the different sheets is clearly visible.
In the elastic region, the physical sheet is adjacent to the (21) sheet, but directly connects to the (22) one above the second threshold.

Although the $\phi$-map correctly unfolds the elastic and inelastic branch cuts, it cannot be directly used in a Taylor series because of the LHC on the (21) sheet.
This LHC starts at $\phi_L \equiv \phi_{(21)}(s_-) < 0$ and ends at $\phi = -1$, and would, thus, limit the region of convergence of such a series to $|\phi| < -\phi_L$.
Similar to before, we therefore need to find a way to unfold the LHC, although we now have the advantage that both the (11) and (21) sheets are within the disc already.

\subsection{The final parametrisation}
\begin{figure}[t]
    \centering
    \begin{subfigure}{0.48\textwidth}
        \includegraphics[width=\textwidth]{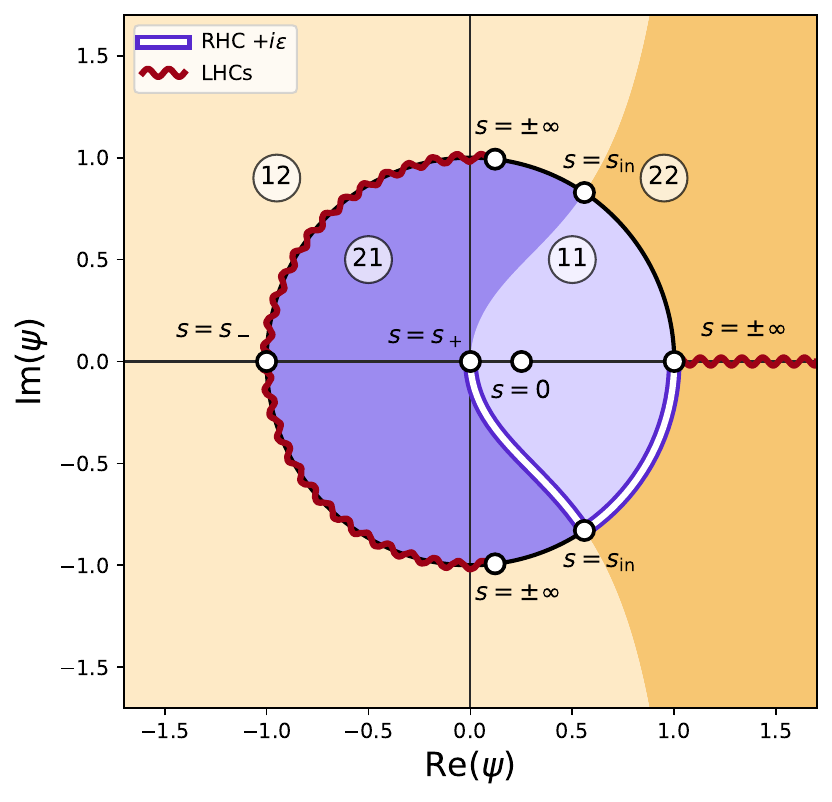}
    \end{subfigure}
    \begin{subfigure}{0.48\textwidth}
        \includegraphics[width=\textwidth]{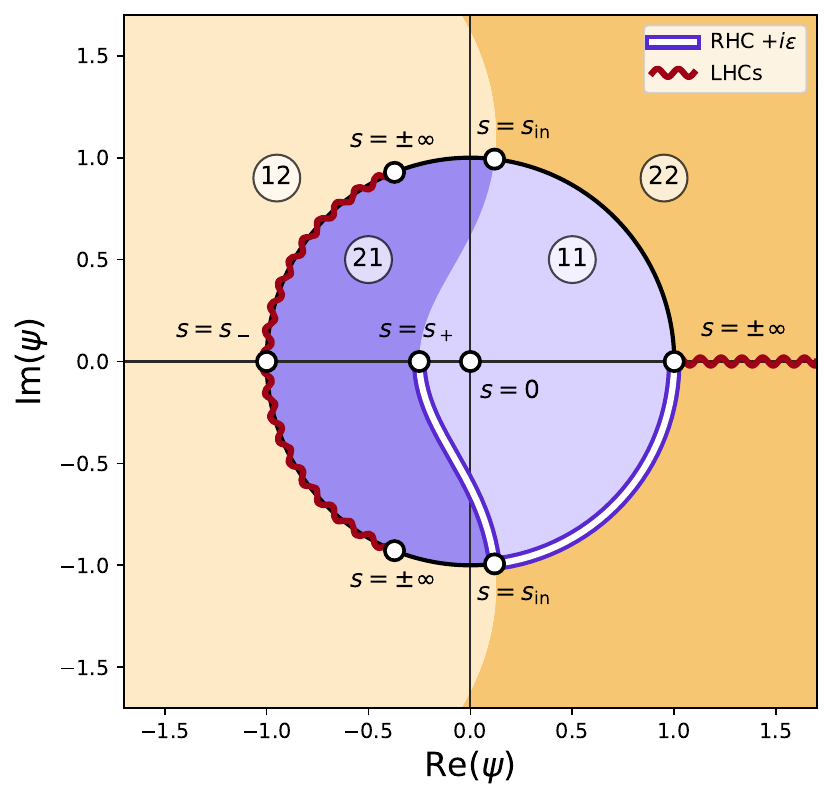}
    \end{subfigure}
    \caption{\label{fig:psi_map}
        The complex $\psi$-plane with $s_0 = s_+$ (left) and $s_0 = 0$ (right).
        The LHC of the (21) sheet is pushed on the unit circle.
        In contrast to the $\zeta$-map, this cut is not opened, and still presents a discontinuity in the $\psi$-plane.
        Higher-energy branch cuts are mapped to the arc of the unit circle surrounding $\psi = 1$, and are not represented.
    }
\end{figure}

As for the $\zeta$-map, we can circumvent the left-hand-cut limitation by pushing the cut onto the unit circle itself.
This involves distorting some of the lines that separate the Riemann sheets in the $\phi$-plane, but it can be done while keeping the rest of the (11) and (21) Riemann sheets inside the unit disc.
This is achieved by using another map, called $\chi$, which is derived in more detail in \cref{app:lhc_push}.
It is given by
\begin{align} \label{eq:chi_map}
    \chi(x,x_L,x_0)\equiv\frac{\sqrt{A}-\sqrt{B}}{\sqrt{A}+\sqrt{B}}\,, \quad\mathrm{with~}\begin{cases}
        A = \bigl(x(x_L-1)^2-x_L(x-1)^2\bigr)(x_0-1)^2 \\
        B = \bigl(x_0(x_L-1)^2-x_L(x_0-1)^2\bigr)(x-1)^2
    \end{cases},
\end{align}
where $x_L$ is the point at which the left-hand cut opens, that is pushed onto the border of the unit disc, and $x_0$ is a free parameter that satisfies $\chi(x_0, x_L, x_0) = 0$.
For clarity, we restricted the definition of $\chi$ to the region $|x| \leq 1$ in \cref{eq:chi_map}.
To avoid any unexpected behaviour, we therefore use a piece-wise definition for our final 4-sheet map, which we call $\psi$ and write as
\begin{align}\label{eq:psi_map}
   \psi(s;s_+,s_\mathrm{in},s_-,s_0)=
   \begin{cases}
       \chi(\phi(s;s_+,s_\mathrm{in}),\phi_L,\phi_0) \quad&\text{for}\,|\phi|\leq1\\[5pt]
       \dfrac{1}{\chi\left(\frac{1}{\phi(s;s_+,s_\mathrm{in})},\phi_L,\phi_0\right)} \quad&\text{for}\,|\phi|>1  
   \end{cases}\,,
\end{align}
where $\phi$ is defined in \cref{eq:phi_map}.
Above, $\phi_L \equiv \phi_{(21)}(s_-)$ is the branch point of the left-hand cut in $\phi$, and $\phi_0 \equiv \phi_{(11)}(s_0)$ corresponds to the point mapped to $\psi_{(11)}(s_0) = 0$.
Analogously, the inverse of $\psi$ is given by 
\begin{align}\label{eq:psi_map_inv}
   \phi(\psi;s_+,s_\mathrm{in},s_-,s_0)=
   \begin{cases}
       \chi^{-1}(\psi,\phi_L,\phi_0) \quad&\text{for}\,|\psi|\leq1\\[5pt]
       \dfrac{1}{\chi^{-1}\left(\frac{1}{\psi},\phi_L,\phi_0\right)} \quad&\text{for}\,|\psi|>1  
   \end{cases}\,,
\end{align}
with
\begin{align} \label{eq:chi_map_inv}
    \chi^{-1}(y,x_L,x_0)\equiv\frac{\sqrt{A+B}-\sqrt{A}}{\sqrt{A+B}+\sqrt{A}}\,, \quad\mathrm{with~}\begin{cases}
        A = (y-1)^2 (x_L-1)^2 (x_0-1)^2 \\
        B = 4 x_0 (y+1)^2 (x_L-1)^2 \\
        \phantom{B=}- 16 y x_L (x_0-1)^2
    \end{cases}.
\end{align}
The complex $\psi$-plane with the left-hand cut pushed to the boundary of the unit circle is pictured in \cref{fig:psi_map},
for the choices $s_0 = s_+$ and $s_0 = 0$.
In contrast to the $\zeta$-map, which aims at opening the LHC on the second sheet, the $\psi$-variable still presents a cut when crossing the arc of the circle.
In \cref{fig:psi_map}, we kept the (12) label for the sheet behind this cut; however, our parametrisation does not differentiate this sheet from the $(21^-)$ sheet obtained by analytically continuing the $(21)$ sheet across the cut.

The visible distortion of the boundaries between the sheets and the size of the arc covered by the LHC increases with the distance between the elastic and inelastic thresholds, $s_+$ and $s_\mathrm{in}$.
In the limit $s_\mathrm{in} \to \infty$, the $\psi$-map reduces to the $\zeta$-map
\begin{equation}\label{eq:psi_limit}
    \lim_{s_\mathrm{in} \to \infty} \psi(s;s_+,s_\mathrm{in},s_-,s_0) = \zeta(s;s_+,z_0=z(s_0,s_+,s_-))\,.
\end{equation}

Using the $\psi$-map, we can finally parametrise the form factors using the SSE
\begin{equation}
    F(s) = \mathcal{P}(\psi) \sum_i a_i \psi^i ~ \Big|_{\psi = \psi(s)} \,,
\end{equation}
where $\mathcal{P}(\psi)$ is an appropriate product of poles that implements bound states and resonances below or above the inelastic threshold.
This parametrisation can be easily refined by imposing the correct threshold behaviour since it currently implements a square-root behaviour at the thresholds $s_+$ and $s_\text{in}$.
Finally, the known high-energy behaviour of the form factor can be recovered if the parametrisation is modified by a prefactor obtained from the relation
\begin{equation}
    \left(\psi(s)-1\right)^2 \underset{s\to\infty}{\sim} \frac{1}{s},
\end{equation}
or by imposing a constraint on the coefficients.

\begin{table}[t]
    \centering
    \begin{tabular}{ccccc}
        \toprule
        resonance $r$ & sheet $S$ & Ref. & $\sqrt{s}_{r,S}\,[\mathrm{GeV}]$ & $\psi_{r,S}$ \\
        \midrule
        $f_0(500)$ & $(21)$ & \cite{Garcia-Martin:2011nna} & $0.457-0.279i$ & $-0.349-0.537i$ \\
        $f_0(980)$ & $(21)$ & \cite{Garcia-Martin:2011nna} & $0.996-0.025i$ & $-0.003-0.927i$ \\
        $f_0(980)$ & $(22)$ & \cite{Mao:2009cc} & $0.977-0.060i$ & $\phantom{-}0.328-1.179i$ \\
        \bottomrule
    \end{tabular}
    \caption{Comparison between the literature pole positions of the $f_0(500)$ and $f_0(980)$ resonances in the conformal $\psi$-plane, $\psi_{r,S}=\psi_{(S)}\bigl(s_{r,S},s_+=4M_\pi^2,s_\mathrm{in}=4M_K^2,s_-=0,s_0=0\bigr)$.}
    \label{tab:polepositions_CC}
\end{table}

\subsection{Application to the scalar pion form factor}
We now test the $\psi$ SSE using pseudo-data again.
The $s_\mathrm{in} \to\infty$ limit in \cref{eq:psi_limit} already ensures that the $\psi$ SSE performs at least as well as the $\zeta$ one on the test cases of \cref{sec:application_SingleChannel}, i.e., on models that do not implement an inelastic threshold.
To further test the possibilities of the $\psi$ SSE, we study a more realistic example of the scalar pion form factor.
Since we can now parametrise the form factor above the elastic threshold, we simultaneously fit the strange and non-strange form factors, as defined in Refs.~\cite{Donoghue:1990xh,Moussallam:1999aq,Descotes-Genon:2000byu,Hoferichter:2012wf,Daub:2012mu,Celis:2013xja,Daub:2015xja,Winkler:2018qyg}.
We implement the resonances up to the $f_0(980)$, and the product of poles takes the same form for both form factors
\begin{equation}
    \mathcal{P}(\psi) = \prod_{r=1}^3 \frac{1}{(\psi-\psi_r)(\psi-\psi_r^*)},
\end{equation}
with one $f_0(500)$ pole $\psi_1=\psi_{(21)}\bigl(s_{f_0(500)}\bigr)$ on the $(21)$-sheet and two $f_0(980)$ poles $\psi_2=\psi_{(21)}\bigl(s_{f_0(980),(21)}\bigr)$ and $\psi_3=\psi_{(22)}\bigl(s_{f_0(980),(22)}\bigr)$ on the $(21)$-sheet and the $(22)$-sheet, respectively.
As detailed in Ref.~\cite{Burkert:2022bqo}, the position of the resonance poles on the Riemann sheets that are not contiguous to the physical sheet does not depend on the resonance mass and width only.
We treat, therefore, all these poles as uncorrelated.
We provide the position of these poles in the $\psi$-plane, as found in the literature, in \cref{tab:polepositions_CC}.
Although the $f_0(500)$ and $f_0(980)$ are expected to have poles on all the unphysical sheets, we only implement the ones listed above, as they are the closest to our input data and, therefore, the only ones relevant to our simple analysis.
As for the previous tests of \cref{sec:application_SingleChannel}, we do not impose any additional constraints on the high-energy or threshold behaviour of the SSE.
Consequently, our fit parameters are, again, the shared pole locations $\sqrt{s_r}$, $r=1,2,3$, and the two independent sets of coefficients for the SSEs of the non-strange and strange pion FFs.

\begin{table}[t]
    \centering
    \begin{tabular}{ccccc}
        \toprule
        $n$ & $\chi^2\,[10^{-3}]$ & $\sqrt{s}_{f_0(500)}$ [GeV] & $\sqrt{s}_{f_0(980),(21)}$ [GeV] & $\sqrt{s}_{f_0(980),(22)}$ [GeV] \\
        \midrule
        Refs.~\cite{Garcia-Martin:2011nna,Mao:2009cc} & -- & $0.457-0.279i$ & $0.996-0.025i$ & $0.977-0.060i$ \\
        \midrule
        7 & $533$ & $0.418-0.265i$ & $1.002-0.028i$ & $0.966-0.082i$ \\
        9 & $158$ & $0.448-0.317i$ & $1.000-0.031i$ & $0.926-0.131i$ \\
        11 & $72.3$ & $0.450-0.275i$ & $1.001-0.032i$ & $0.983-0.037i$ \\
        \bottomrule
    \end{tabular}
    \caption{\label{tab:Fits_SFF_CoupledChannel}
        Extracted pole positions for the $f_0(500)$ and $f_0(980)$ from fits in the variable $\psi$ to the coupled-channel $I=0$, $l=0$ form factors.
        For comparison we list the $(21)$-sheet pole positions for the $f_0(500)$ and $f_0(980)$ from Ref.~\cite{Garcia-Martin:2011nna}, as well as the $(22)$-sheet pole position for the $f_0(980)$ from Ref.~\cite{Mao:2009cc}.
        The quoted $\chi^2$ is computed by assuming uncorrelated uncertainties of 1 for all 636 input data points.
        For $n \geq 11$, the fit showed several nearly degenerate local minima, indicating the need for a more comprehensive statistical analysis.
    }
\end{table}

The input pseudo-data is generated by sampling 636 data points in the range $s\in[0,1.5]\,\mathrm{GeV}^2$ from the coupled-channel Omnès matrix of Ref.~\cite{Daub:2015xja}.\footnote{%
The large number of data points directly stems from the input data file of the authors of Ref.~\cite{Daub:2015xja}.
Reducing the number of input points by a factor of 10 leads to a decrease of the $\chi^2$ from 0.0723 to 0.0015 for 606 and 34 degrees of freedom, respectively.
Doing so, the pole parameters are only altered at the percent level.}
Our fit results for different truncation orders are provided in \cref{tab:Fits_SFF_CoupledChannel}, and the shape of the form factors for $n = 11$ are shown in \cref{fig:SFF_CC_Fits}.
The best-fit parameters are also provided in the supplementary material.
We find that a higher truncation order is required than in the previous analysis to obtain a visually satisfactory description of the data.
This is because this parametrisation aims to describe the first two sheets completely.
Yet, with $n = 11$, we obtain a visually excellent description of the data in the region where convergence is expected, namely below $s = 1.5\,\mathrm{GeV}^2$, where our input model breaks down.
For larger values of $n$, the fit exhibits many local minima, so a more in-depth statistical analysis would be needed.
We find that the parametrisation is able to reproduce the position of the resonance poles, even on the $(21)$ and $(22)$ sheets, albeit with errors.
As mentioned above and discussed in the literature, the position of the $f_0(980)$ poles on these two sheets differs~\cite{Burkert:2022bqo}.
\begin{figure}[t]
    \centering
    \begin{subfigure}{0.49\textwidth}
        \includegraphics[width=\textwidth]{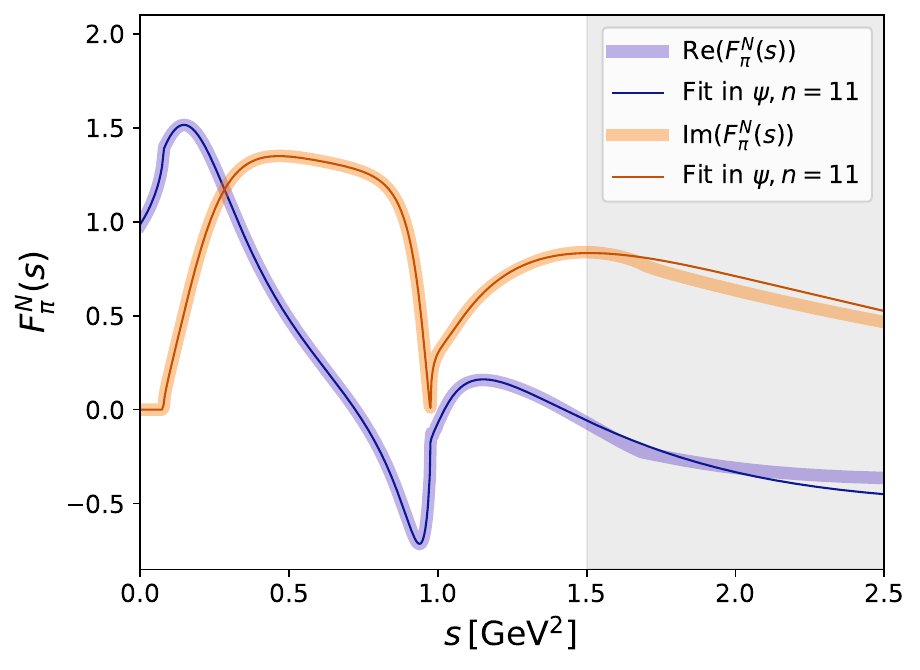}
    \end{subfigure}
    \begin{subfigure}{0.49\textwidth}
        \includegraphics[width=\textwidth]{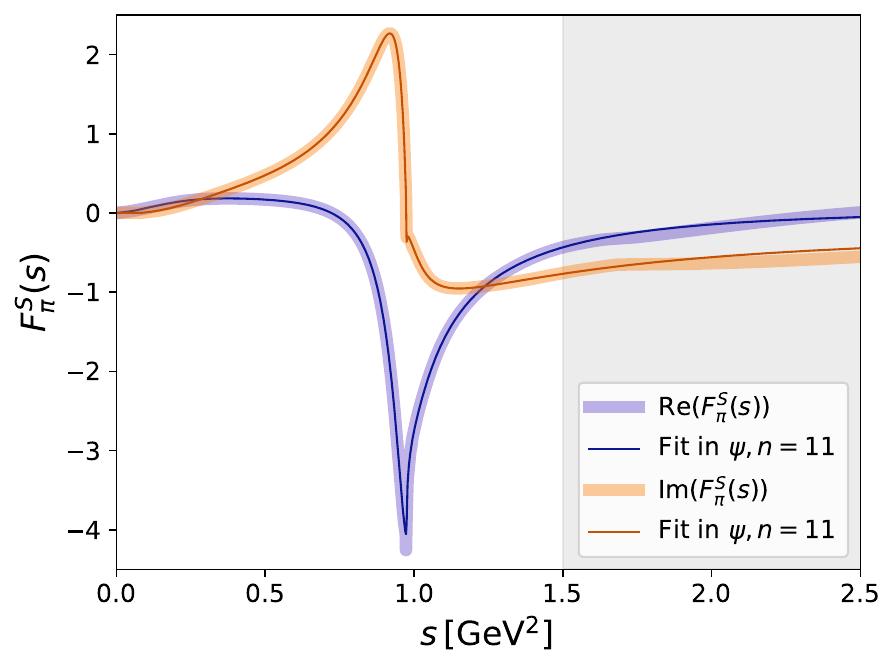}
    \end{subfigure}
    \caption{Coupled-channel fits in $\psi$ for the non-strange and strange scalar pion form factors $F_\pi^N$ and $F_\pi^S$ for $n=11$.
    The fit range is $s \in [0,1.5]\,\mathrm{GeV}^2$, the grey area shows where the model has been continued to higher energies and is therefore not fitted.}
    \label{fig:SFF_CC_Fits}
\end{figure}

\section{Conclusions \texorpdfstring{\&}{and} outlook}
\label{sec:conclusions}
The two novel parametrisations introduced here allow for a more detailed study of the second Riemann sheet of hadronic form factors, including resonance poles, as well as the left-hand cut required by unitarity.
The first parametrisation is ideal to study purely theoretical form factors with only a single channel, such as the pion vector and scalar form factors obtained from the mIAM.
The second parametrisation explicitly accounts for the second two-particle threshold, meaning it can be used in the future to analyse experimental data and extract phase shifts, resonance poles, and residues.
Sample applications improved the convergence properties of both parametrisations with respect to the commonly used $z$-expansion, demonstrating the advantage of directly encoding as many of the analytical structures of form factors as possible in the expansion variable.

A direct application of the conformal four-sheet map would be a fit to $e^+e^-\to\pi^+\pi^-$ and $\tau^- \rightarrow \pi^-\pi^0\nu_\tau$ experimental data to extract the resonances' pole locations and residues.
In contrast to the $z$-expansion approach of Ref.~\cite{Kirk:2024oyl}, the $\rho(770)$ pole is directly within the convergence radius, allowing for a more detailed study of the pole properties.

Several variations of the parametrisations are also of interest.
A generalisation to the case with a first channel that features unequal-mass particles would allow for studies of $\tau \rightarrow K\pi\nu_\tau$ decays and applications to semileptonic decays, for example, $D\rightarrow \pi\ell\nu_\ell$ decays where the $D^\ast$ resonance lies closely above the first threshold.
In addition to the high-energy behaviour, also the behaviour at $s_+$, $s_\text{in}$, and $s_-$ of form factors is known and, thus, can be implemented in the parametrisation, either following Refs.~\cite{Bourrely:2008za,Herren:2025cwv} by expanding in suitable polynomials, or by directly imposing constraints on the expansion coefficients.
Going beyond mesons, our method to incorporate resonance poles in a way that respects all analytic properties of partial waves and production amplitudes may prove useful also for the precise extraction of resonance properties in meson--baryon systems, where methods going beyond the Breit--Wigner approximation are urgently sought after; cf., e.g., Refs.~\cite{Svarc:2013laa,Svarc:2022buh}.
Incorporating the four-sheet map in the Khuri--Treiman formalism would allow us to improve the parametrisation of $B\rightarrow\pi\pi\ell\nu$ decays introduced in Ref.~\cite{Herren:2025cwv}.
Finally, going beyond two-particle thresholds or two channels is phenomenologically significant, but it involves function spaces beyond those considered here.
However, first investigations of the three-channel problem, albeit without taking into account left-hand cuts, have been conducted in Ref.~\cite{Yamada:2022xam}.

\section*{Acknowledgements}  
We thank Christoph Hanhart for useful discussions.
This research was supported in part by the Swiss National Science Foundation (SNF) under contract 200021-212729, and by the MKW NRW under the funding code NW21-024-A.

\appendix

\section{Riemann sheet structure from two-particle intermediate states} \label{app:riemann_sheets}

Let us briefly review the Riemann sheet structure arising from a two-particle intermediate state by following the example of the scalar one-loop function
\begin{align}
    B_0(s) \equiv B_0\bigl(q^2=s;m_1^2,m_2^2\bigr) = \frac{1}{i\pi^2} \int \d^4\ell\,\frac{1}{(\ell^2-m_1^2+i\epsilon)((\ell-q)^2-m_2^2+i\epsilon)}\,.
\end{align}
It is analytic in the complex variable $s$ up to singular points that can be determined by the means of Landau's equations~\cite{Landau:1959fi}.
Here, they take the form
\begin{align}
    0 &= \alpha_1 \, \ell + \alpha_2 \, (\ell-q)\,, \notag\\
    0 &= \alpha_1 \, (\ell^2-m_1^2) = \alpha_2 \, ((\ell-q)^2-m_2^2)\,,
\end{align}
with the Feynman parameters summing to one, $\alpha_1+\alpha_2=1$.
Its leading solutions, i.e., solutions with $\alpha_i \neq 0$, stemming from both internal particles going on-shell, are
\begin{align}
    s_+ &= (m_1+m_2)^2, \qquad\text{with } \alpha_1 = \frac{m_2}{m_1+m_2}\,,\quad \alpha_2=\frac{m_1}{m_1+m_2}\,, \notag\\
    s_- &= (m_1-m_2)^2, \qquad\text{with } \alpha_1 = \frac{m_2}{m_2-m_1}\,,\quad \alpha_2=\frac{m_1}{m_1-m_2}\,.
\end{align}
In addition to their positions, Landau's analysis also predicts the types of the resulting singularities.
In this case, they are square-root branch points.
The solutions for $\alpha_1$ and $\alpha_2$ corresponding to $s_+$ are real and positive, lying within the Feynman parameter integration range.
This causes a branch cut in $s$ on both the first Riemann sheet of $B_0(s)$ and all other sheets that can be reached by deforming the integration contour.
In contrast, this is not the case for $s_-$, as one of the corresponding $\alpha_i$ is negative, which only leads to singularities on the other Riemann sheets.

The resulting topological structure of the Riemann surface of $B_0(s)$ looks as follows.
On the first Riemann sheet $B_{0,\mathrm{I}}(s)$, there is only the square-root-type right-hand cut (RHC), which runs from $s_+$ towards $+\infty$.
Analytic continuation through this cut leads to a second Riemann sheet,\footnote{%
Here, we use the naming convention from Ref.~\cite{Jing:2025qmi}, where an explicit visualisation of this topology can be found in Fig.~2.
} $B_{0,\mathrm{II}}(s) \equiv B_{0,\mathrm{II}^+}(s)$, which has the same RHC as the first sheet and an additional left-hand cut (LHC) extending from $s_-$ to $-\infty$.
Crossing this LHC leads to another sheet, $B_{0,\mathrm{II}^-}(s)$, which in turn features a new RHC.
Alternating between the RHCs and LHCs eventually leads to an infinite cascade of Riemann sheets, $B_{0,N^\pm}(s)$, $N=\mathrm{II},\,\mathrm{III},\,\ldots$, all connected through these cuts and meeting at complex infinity.
For this simple example of the scalar loop function, one can explicitly write down the analytic continuation to all these sheets as
\begin{align}
    B_{0,N^\pm}(s) = B_0(s) \pm (N-1) \, l(s) \qquad \text{with } l(s) = \frac{2\pi}{s} \sqrt{s_+-s}\sqrt{s-s_-}\,,
\end{align}
which explains the naming scheme.
The function $l(s)$ can be obtained, for example, via Cutkosky rules or via the discontinuity calculus developed in Ref.~\cite{Jing:2025qmi}.

In general, while the exact details of the analytic continuation may differ, any two-particle intermediate state results in the same two thresholds $s_+$ and $s_-$ and the same infinite cascade of Riemann sheets.
Rather than simple square-root cusp behaviour at the thresholds, the $l$-th partial wave of the two-particle system opens as $(s-s_\pm)^{(2l+1)/2}$.
However, this does not change the square-root nature of the branch points (as opposed to logarithmic-type branch points, which split into infinitely many sheets).

\section{One-threshold \texorpdfstring{$\boldsymbol\zeta$}{ζ}-variable with \texorpdfstring{$\boldsymbol{2^n}$}{2ⁿ} sheets} \label{app:zeta}

In \cref{eq:zeta_map_definition}, we introduce a new conformal variable $\zeta(s)$ that maps the first and second Riemann sheets of a hypothetical one-threshold form factor into the unit disc.
This is achieved by unfolding the left-hand cut of the second sheet of the standard $z$-plane in \cref{fig:zmap} onto the unit circle.
Similarly to how the original $z$-variable features the second sheet as a ``reflection'' at the unit circle, $z_{\mathrm{II}}=1/z_{\mathrm{I}}$,
this new map also has two more ``reflected'' sheets outside of the unit circle, $\zeta_{\mathrm{II^-}}=1/\zeta_{\mathrm{II}}$ and $\zeta_{\mathrm{III^+}}=1/\zeta_{\mathrm{I}}$, as can be seen in \cref{fig:zeta_map}.
In the complex $\zeta$-plane, these four Riemann sheets are nested around each other.
As discussed in \cref{app:riemann_sheets}, the outermost layer $F_{\mathrm{III^+}}(s)$ features another LHC, which in the $\zeta$-plane runs from $\zeta=1$ to $\zeta=\infty$ if $z_0=0$ is used.

In principle, one can repeat the same procedure by iteratively unfolding the appearing LHCs.
This would result in doubling the amount of layered Riemann sheets mapped into the unit disc each time.
Denoting with $n$ the number of times the $\omega$-map defined in \cref{eq:cmap_definition} is applied to unfold LHCs, this yields the following series of recursively defined new variables,
\begin{align}
    \zeta^{(1)}(s;s_+,\zeta_0) &\equiv \zeta(s;s_+,z_0=\zeta_0) = \omega\bigl(z(s;s_+,s_-),\zeta_0\bigr)\,, \notag\\
    \zeta^{(n+1)}(s;s_+,\zeta_0) &\equiv \omega\bigl(\zeta^{(n)}(s;s_+,0);\zeta_0\bigr)\,.
\end{align}
For each of these variables, $\zeta_0$ is the only free parameter and determines the point that is mapped to the origin.
For a fixed number $n$, the map $\zeta^{(n)}$ features $2^n$ sheets within the unit disc and behaves as
\begin{align}
    \bigl(\zeta^{(n)}(s)-1\bigr)^{2(n+1)} \underset{s\to\infty}{\sim} \frac{1}{s}
\end{align}
at high energies.

\section{Theory models used to test the parametrisations} \label{app:th_models}

This section briefly reviews the models used to produce pseudo-data to test the new parametrisations.

\paragraph{Scalar isoscalar pion form factor.}
The NLO results from Chiral Perturbation Theory (ChPT) for the $I=0$, $l=0$ partial wave are taken from Ref.~\cite{Niehus:2020gmf} and read
\begin{align}
    t_2(s) &=\frac{2s-M_\pi^2}{32\pi F^2}\,, \notag\\
    t_4(s) &= C_0^{00} + C_1^{00} L + C_2^{00} L^2 +C_{l_1}^{00}l^r_1 + C_{l_2}^{00}l^r_2 + C_{l_3}^{00} l^r_3 +i\sigma_\pi(s) t_2(s)^2\,,
\end{align}
with
\begin{equation}
    L=\log\frac{1+\sigma_\pi(s)}{1-\sigma_\pi(s)}
\end{equation}
and
\begin{align}
    C_0^{00} &= \frac{373 M_\pi^4 - 190 M_\pi^2 s + 51 s^2 - 5 \left(31 M_\pi^4 - 32 M_\pi^2 s + 10 s^2\right) \log \left( \frac{M_\pi^2}{\mu^2} \right)}{9216 F^4 \pi^3}\,, \notag\\
    C_1^{00} &= \frac{36 M_\pi^6 - 303 M_\pi^4 s + 260 M_\pi^2 s^2 - 50 s^3}{9216 F^4 \pi^3 s \sigma}\,, \notag\\
    C_2^{00} &= \frac{M_\pi^4 (25 M_\pi^2 - 6 s)}{1536 F^4 \pi^3 (4 M_\pi^2 - s)}\,, \quad C_{l_2}^{00} = \frac{28 M_\pi^4 - 20 M_\pi^2 s + 7 s^2}{48 F^4 \pi}\,, \notag\\
    C_{l_1}^{00} &= \frac{44 M_\pi^4 - 40 M_\pi^2 s + 11 s^2}{48 F^4 \pi}\,, \quad C_{l_3}^{00} = \frac{5 M_\pi^4}{16 F^4 \pi}\,.
\end{align}
We use the values $F=88.27\,\mathrm{MeV}$ and $M_\pi=139.57\,\mathrm{MeV}$, while the low-energy constants (LECs) $l^r_i$ are taken from Ref.~\cite{Pelaez:2010fj}, yielding $l_1^r=-3.7\cdot10^{-3}$, $l_2^r=5\cdot10^{-3}$, and $l_3^r=0.8\cdot10^{-3}$.
They are evaluated at the renormalisation scale $\mu=770\,\mathrm{MeV}$.

\begin{figure}[t]
    \centering
    \includegraphics[width=0.48\textwidth]{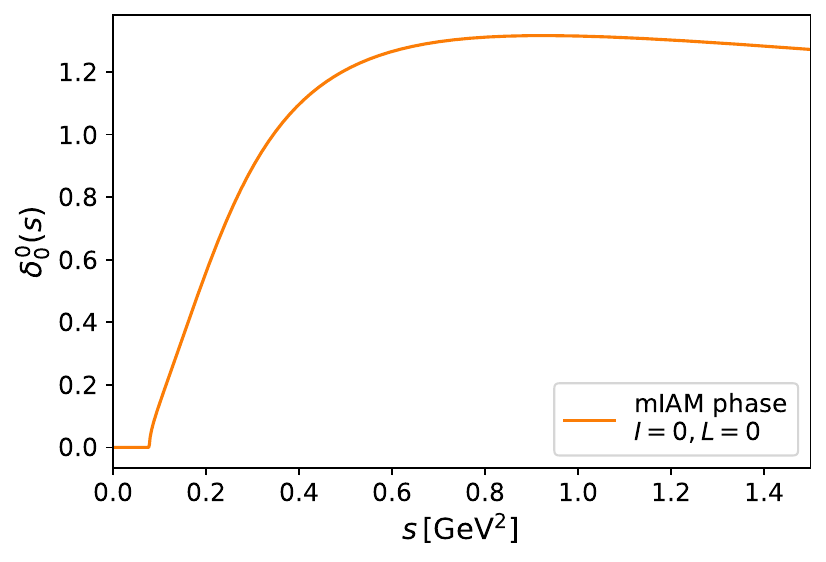}
    \includegraphics[width=0.48\textwidth]{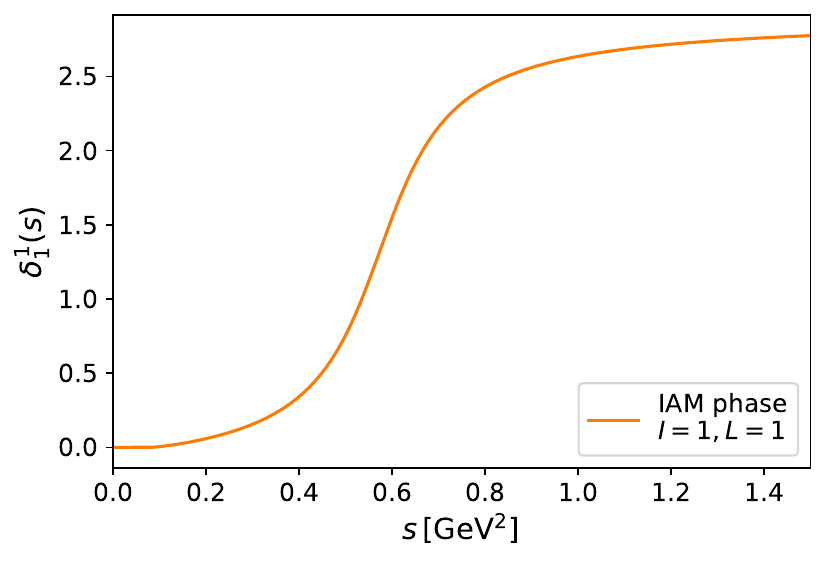}
    \caption{Phases used as input for the single-channel pion scalar and vector form factors, obtained via the (modified) IAM.}
    \label{fig:Phases_IAM}
\end{figure}

Unfortunately, the NLO IAM by itself, which is defined by 
\begin{equation}
    t_\text{IAM}(s)=\frac{t_2^2(s)}{t_2(s)-t_4(s)}\,,
\end{equation}
cannot reproduce the location of the Adler zero $s_A$ correctly, which appears below threshold in scalar waves~\cite{GomezNicola:2007qj}.
Furthermore, it produces spurious poles close to the Adler zero on the real axis on both the first and the second Riemann sheets~\cite{Pelaez:2010fj}.
These issues can be fixed if, instead of the plain IAM, we use a modified version (mIAM), whose derivation can be found in Ref.~\cite{GomezNicola:2007qj}.
We obtain
\begin{equation}
    t_\text{mIAM}(s)=\frac{t_2(s)^2}{t_2(s)-t_4(s)+A_\text{mIAM}(s)}\,,
\end{equation}
with the additional term\footnote{%
This is already the simplified expression for $\pi\pi$ scattering, a more general form can be found in, e.g., Ref.~\cite{Colangelo:2017fiz}.
} 
\begin{equation}
    A_\text{mIAM}(s)=t_4(s_2)-\frac{(s_2-s_A)(s-s_2)[t_2'(s_2)-t_4'(s_2)]}{s-s_A}\,,
\end{equation}
where the Adler zero is set to its NLO approximation of $s_A=s_2+s_4$ and $t_i'(s)$ denotes the derivative of the ChPT partial waves with respect to $s$.
The values for $s_2$ and $s_4$ we use in the numerical calculation are given in Ref.~\cite{Heuser:2024biq} as
\begin{align}
    s_2&=\frac{M_\pi^2}{2}\,, \qquad
    s_4=-\frac{M_\pi^4}{(48\pi F)^2}\Bigr[1163+2(107\bar{l}_1+158\bar{l}_2-90\bar{l}_3)-908A-4224A^2\Bigl]\,,
\end{align}
with $A = \frac{\arctan\sqrt{7}}{\sqrt{7}}$ and the values for the LECs $\bar{l}_1=-2.3$, $\bar{l}_2=6$, and $\bar{l}_3=2.9$ from Ref.~\cite{Gasser:1983yg}.\footnote{%
In Ref.~\cite{Gasser:1983yg}, one can also find a relation between the LECs $l^r_i$ used for the IAM and the $\bar{l}_i$ used here.
}
The phase of the $I=0$, $l=0$ mIAM is pictured in \cref{fig:Phases_IAM} (left).

\paragraph{Vector isovector pion form factor.}
The NLO ChPT results $t_2(s)$ and $t_4(s)$ for the $I=1$, $l=1$ case are given in, e.g., Ref.~\cite{Dax:2018rvs} and read
\begin{align}
    t_2(s) &= \frac{s \sigma_\pi(s)^2}{96 \pi F^2}\,, \notag \\
    t_4(s) &= \frac{t_2(s)}{48 \pi^2 F^2} \biggl[s \left(\bar{l} + \frac{1}{3}\right) - \frac{15}{2} M_\pi^2 - \frac{M_\pi^4}{2s} \Bigl(41 - 2L_{\sigma_\pi}\left(73 - 25\sigma_\pi(s)^2\right) \notag \\
    & + 3L_{\sigma_\pi}^2 \left(5 - 32\sigma_\pi(s)^2 + 3\sigma_\pi(s)^4\right) \Bigr) \biggr]+i\sigma_\pi(s) t_2(s)^2\,,
\end{align}
where
\begin{equation}
    L_{\sigma_\pi} = \frac{1}{\sigma_\pi(s)^2}\left(\frac{1}{2\sigma_\pi(s)}\log\frac{1+\sigma_\pi(s)}{1-\sigma_\pi(s)}-1\right) \,.
\end{equation}
The values of $F=88.27\,\mathrm{MeV}$ and $\bar{l}=48\pi^2(l_2^r-2l_1^r)=5.98$ are taken from Ref.~\cite{Niehus:2020gmf}, and we use the charged pion mass $M_\pi=139.57\,\mathrm{MeV}$.
The resulting phase of the IAM is pictured in \cref{fig:Phases_IAM} (right).

\section{Relation between pole parameters and coupling constants} \label{app:pole_parameters}

As shown in Ref.~\cite{Kirk:2024oyl}, one great advantage of the parametrisations based on a conformal expansion is that they provide direct access to all the couplings of the resonances.
In the vicinity of the $f_0(500)$ and $\rho(770)$ resonance poles, we parametrise the $\pi\pi$ partial-wave amplitudes and form factors on their respective second sheets in the following way,
\begin{align}
    t_{0,\mathrm{II}}^{0}(s) &\sim \frac{g_{\sigma\pi\pi}^2}{16\pi} \, \frac{1}{s_{\sigma}-s}\,, \notag\\
    t_{1,\mathrm{II}}^{1}(s) &\sim \frac{g_{\rho\pi\pi}^2}{48\pi}  \,\frac{s-4M_\pi^2}{s_{\rho}-s}\,,
\end{align}
and
\begin{align}
    F_{\pi,\mathrm{II}}^{S}(s) &\sim g_{\sigma\pi\pi} g_{\sigma q\bar q} \, \frac{1}{s_{\sigma}-s}\,, \notag\\
    F_{\pi,\mathrm{II}}^{V}(s) &\sim \frac{g_{\rho\pi\pi}}{g_{\rho\gamma}} \, \frac{s_{\rho}}{s_{\rho}-s}\,,
\end{align}
to relate the pole residues to the appropriate coupling constants. Here, we used $\sigma\equiv f_0(500)$ for short and denoted by $g_{\sigma q\bar q}$ the coupling of the $f_0(500)$ resonance to a non-strange scalar $q \bar q$ current, similar to Ref.~\cite{Moussallam:2011zg}.
Following the lines of Refs.~\cite{Hoferichter:2017ftn,Hoferichter:2023mgy}, we can also relate the couplings to the form factors evaluated on the first sheet as follows,
\begin{align}
    \frac{g_{\sigma q\bar q}}{g_{\sigma\pi\pi}} &= i \frac{\sigma_\pi(s_\sigma)}{8\pi} F_{\pi,\mathrm{I}}^{S}(s_\sigma) \,, \notag\\
    g_{\rho\pi\pi} g_{\rho\gamma} &= i \frac{\sigma_\pi(s_\rho)^3}{24\pi} F_{\pi,\mathrm{I}}^{V}(s_\rho) \,.
\end{align}
In combination, by computing both the second-sheet residues and the first-sheet values of the form factors at the respective pole locations, we can extract all individual coupling constants simultaneously.

\section{Left-hand-cut transformation} \label{app:lhc_push}

\begin{figure}[t]
    \centering
    \begin{subfigure}{0.75\textwidth}
        \includegraphics[width=\textwidth]{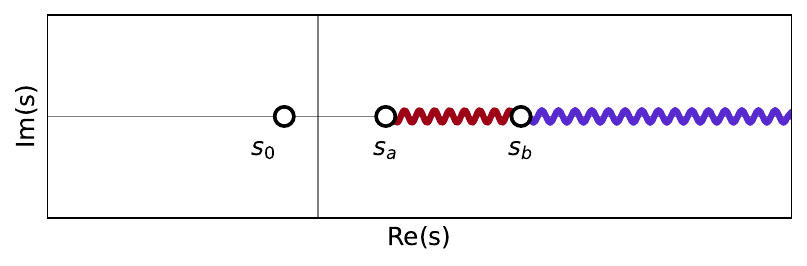}
    \end{subfigure}
    \begin{subfigure}{0.45\textwidth}
        \includegraphics[width=\textwidth]{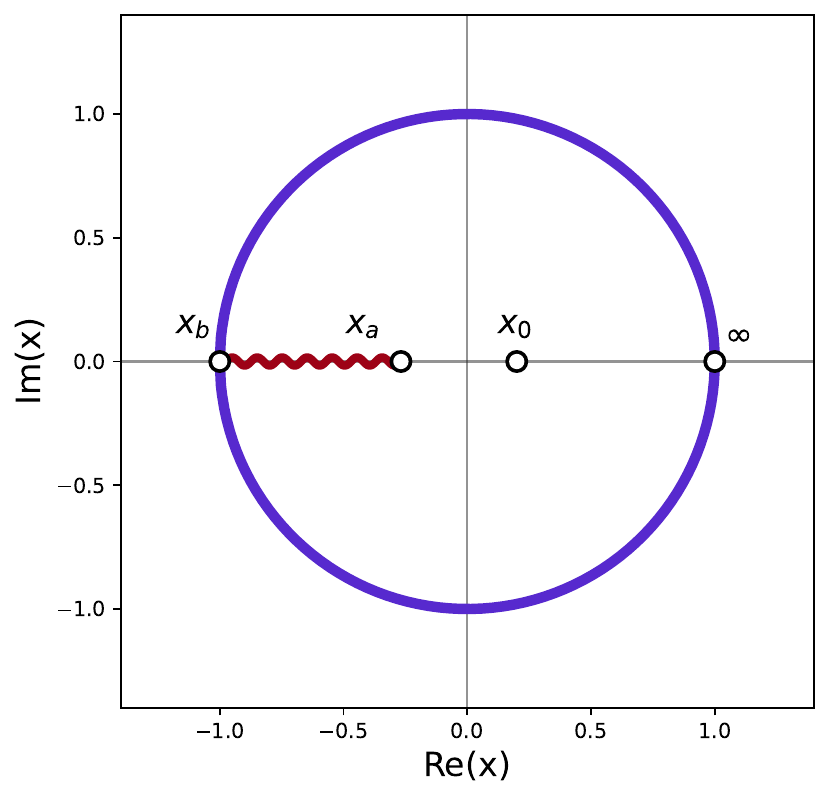}
    \end{subfigure}
    \begin{subfigure}{0.45\textwidth}
        \includegraphics[width=\textwidth]{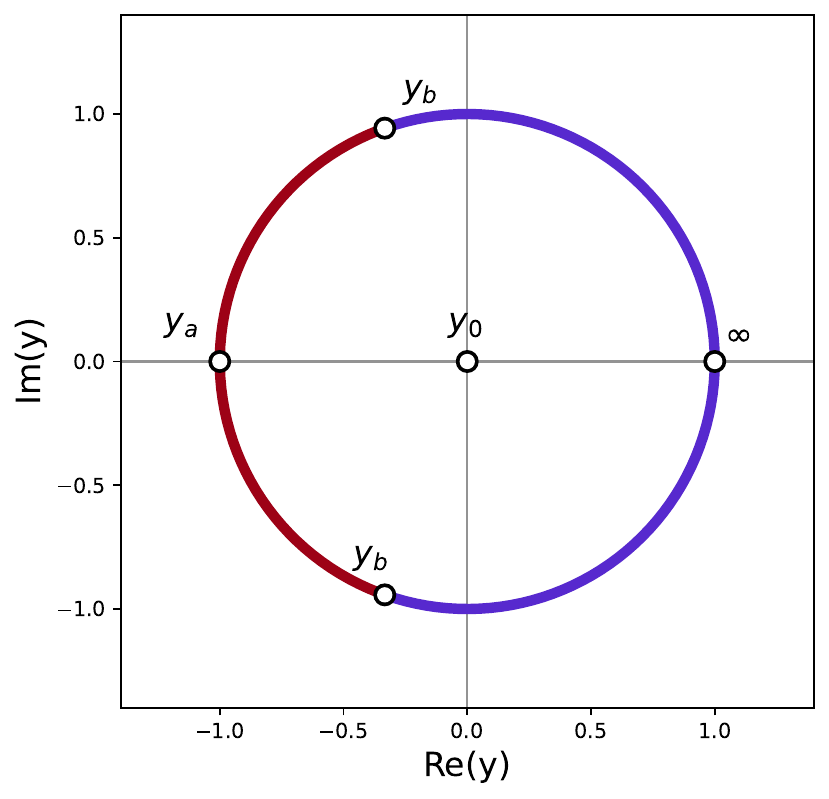}
    \end{subfigure}
    \caption{Top: an unphysical $s$-plane, showing a subthreshold-like cut starting at $s_a$ and a right-hand cut starting at $s_b$. Bottom left: the plane we obtain if we use the map $x=z(s;s_b,0)$ from \cref{eq:unphysical_xys} on the unphysical $s$-plane on top. The points $x_i$ are defined by $x_i=z(s_i;s_b,0)$. Bottom right: the plane we obtain if we use the map $y=z(s;s_a,s_0)$ from \cref{eq:unphysical_xys} on the unphysical $s$-plane on top. The points $y_i$ are defined by $y_i=z(s_i;s_a,s_0)$.}
    \label{fig:chi_map_idea}
\end{figure}

This section describes how the $\chi$ map, which allows us to implement the left-hand cut of the $(21)$ Riemann sheet, is constructed.
The mechanism we apply is inspired by the use of the $z$-map to account for subthreshold cuts in Ref.~\cite{Gopal:2024mgb}.
This reference uses two $z$-maps, which only differ from each other with respect to their thresholds, $s_+$ (both planes are pictured in \cref{fig:chi_map_idea} in the bottom row).
To distinguish the two corresponding $z$-planes, we call the one with the cut inside the unit circle $x$ and the one with the cut on the boundary of the unit circle $y$ in the following.
These maps are defined by
\begin{equation}\label{eq:unphysical_xys}
    x \equiv z(s;s_b,0) \,, \qquad
    y \equiv z(s;s_a,s_0) \,,
\end{equation}
respectively, with $z$ referring to the original $z$-map from \cref{eq:z_map}.
Analogously, the inverse maps are given by $s(x;s_b,0)$ and $s(y;s_a,s_0)$.
The idea of the $\chi$-map is to identify the $\phi$-plane in \cref{fig:phi_map} with the lower-left plane in \cref{fig:chi_map_idea}, and use two consecutive maps to obtain the lower-right plane, which is free of cuts in the disc.

In practice, we use an intermediate, unphysical $s$-plane (top plot in \cref{fig:chi_map_idea}) in which we treat our left-hand cut like a subthreshold cut starting at $s_a$ and ending at $s_b$.
The $\chi$-map from one plane to the other is thus obtained with
\begin{align}
    y &= \chi(x,x_a,x_0) \equiv z\bigl(s(x;s_b,0);s_a,s_0\bigr) = z\bigl(s(x;s_b,0);s(x_a;s_b,0),s(x_0;s_b,0)\bigr)\notag\\
    &=\frac{\sqrt{A}-\sqrt{B}}{\sqrt{A}+\sqrt{B}}\,, \quad\mathrm{with~}\begin{cases}
        A = \bigl(x(x_a-1)^2-x_a(x-1)^2\bigr)(x_0-1)^2 \\
        B = \bigl(x_0(x_a-1)^2-x_a(x_0-1)^2\bigr)(x-1)^2
    \end{cases},
\end{align}
which only depends on the starting point $x_a$ of the cut inside the unit circle in $x$ and the point $x_0$ that we map to the centre of the $y$-plane.
Following the same strategy, the inverse of this transformation, from $y$ back to $x$, can then be written as
\begin{align}
    x &= \chi^{-1}(y,x_a,x_0) \equiv z\bigl( s\bigl(y;s(x_a;s_b,0),s(x_0;s_b,0)\bigr); 1,0 \bigr)\notag\\
    &=\frac{\sqrt{A+B}-\sqrt{A}}{\sqrt{A+B}+\sqrt{A}}\,, \quad\mathrm{with~}\begin{cases}
        A = (y-1)^2 (x_a-1)^2 (x_0-1)^2 \\
        B = 4 x_0 (y+1)^2 (x_a-1)^2 - 16 y x_a (x_0-1)^2
    \end{cases},
\end{align}
which still only depends on $x_a$ and $x_0$.
For our purposes, $x_a$ translates to the point where the left-hand cut opens up on the $(21)$ sheet in $\phi$.
When using $\chi$ to define our resulting map, however, we need to consider that in the process of deriving $\chi$ we used the inverse of the $z$-map $s(z)$, which is not an injective function since $s(z)=s(z_\text{II})=s(1/z)$.
As a consequence, we need to distinguish carefully between input values with an absolute value $\leq 1$ or $>1$ when using $\chi$.
Strictly speaking, it is only valid for input values with an absolute value $\leq 1$ in general, as it will otherwise introduce unwanted cuts in our map, making it necessary to employ a piece-wise definition when using it for a variable that also should be able to take input values $\geq 1$.

One can also express $\chi$ in terms of the map $\omega$ from \cref{eq:cmap_definition} as
\begin{align}
    \chi(x,x_a,x_0) &= -\omega\biggl(\frac{\omega^{-1}(-x,0)}{\omega^{-1}(-x_a,0)},\frac{\omega^{-1}(-x_0,0)}{\omega^{-1}(-x_a,0)}\biggr)\,, \notag\\
    \chi^{-1}(y,x_a,x_0) &= -\omega\biggl(\omega^{-1}(-x_a,0)\cdot\omega^{-1}\biggl(-y,\frac{\omega^{-1}(-x_0,0)}{\omega^{-1}(-x_a,0)}\biggr),0\biggr)\,.
\end{align}

\medskip

\paragraph{Supplementary material and data.}
The following file is provided as supplementary
material/data and is accessible as an ancillary file.
\begin{itemize}
    \item \texttt{fit\_parameters.txt} (Fit Parameters). Examples for the best fit parameters in the following cases:
    (1) $I=0$, $l=0$, $n=5$ fit in $z$; 
(2) $I=0$, $l=0$, $n=5$ fit in $\zeta$; 
(3)~$I=1$, $l=1$, $n=5$ fit in $z$; 
(4) $I=1$, $l=1$, $n=5$ fit in $\zeta$; 
(5) $I=0$, $l=0$, $n=11$ fit in $\psi$.
\end{itemize}

\bibliographystyle{jhep_mod} 
\bibliography{main.bib}

\end{document}